\pdfoutput=1
\documentclass[twocolumn,english,aps,superscriptaddress,prb]{revtex4-1}

\usepackage{babel}
\usepackage{amsmath}
\usepackage{amssymb}
\usepackage{graphicx}

\begin{document}

\title{Comb tensor networks}

\author{Natalia Chepiga}
\affiliation{Department of Physics and Astronomy, University of California, Irvine, CA 92697, USA}
\affiliation{Institute for Theoretical Physics, University of Amsterdam, Science Park 904, 1098 XH Amsterdam, The Netherlands}
\author{Steven R. White}
\affiliation{Department of Physics and Astronomy, University of California, Irvine, CA 92697, USA}

\date{\today}
\begin{abstract} 
In this paper we propose a special type of a tree tensor network 
that has the geometry of a comb---a 1D backbone with finite 1D teeth projecting out from it.
This tensor network is designed to provide an effective description of higher
dimensional objects with special limited interactions, or, alternatively,
one-dimensional systems composed of complicated zero-dimensional objects.  We
provide details on the best numerical procedures for the proposed network, including
an algorithm for variational optimization of the wave-function as a comb
tensor network, and the transformation of the comb into a matrix product state.
We compare the complexity of using a comb versus alternative
matrix product state representations using density matrix renormalization group
(DMRG) algorithms. 
As an application, we study a spin-1 Heisenberg model system which has a
comb geometry. In the case where the ends of the teeth are terminated by spin-1/2 spins, we 
find that Haldane
edge states of the teeth along the backbone 
form a critical spin-1/2 chain, whose properties can
be tuned by the  coupling constant along the
backbone. By adding next-nearest-neighbor interactions along the
backbone, the comb can be brought into a gapped phase with a long-range
dimerization along the backbone. The critical and dimerized phases are
separated by a Kosterlitz-Thouless phase transition, the presence of
which we confirm numerically. Finally, we show that when the teeth contain an odd number of spins
and are not terminated by spin-1/2's, a special type of comb edge states emerge.
\end{abstract}

\maketitle


\section{Introduction}

Modern condensed matter physics to a large extent relies on numerical
simulations. The Density Matrix Renormalization Group (DMRG)\cite{dmrg1,dmrg2}
algorithm has established itself as one of the most powerful numerical tools for
strongly-correlated one-dimensional systems. The method is variational and in
most cases its accuracy is well controlled by the number of
states kept and the number of sweeps---two external parameters of the algorithm.  In
contrast to Wilson's numerical renormalization group approach, the selection of the
effective basis vectors is not based on energies, but on the eigenvalues of the
density matrix, or equivalently, on the entanglement.  The success of DMRG
in one-dimension inspires its further application to two-dimensional systems by
mapping the 2D lattice to a 1D path in a snake-line way.
However, according to the area law\cite{area_law} the entanglement for multiple
leg ladders  grows exponentially with the number of legs. This limit the
width $L$ of these quasi-two dimensional lattices, and the current state of the art
for two-dimensional cylinders with DMRG is about $L\approx 6-14$, depending primarily
on the number of states per site in the model.

The reformulation of conventional DMRG in terms of matrix product states
(MPS)\cite{dmrg3,dmrg4} led to a number of generalizations, most notably for
higher dimensions.  Probably the best
known example is the projected entangled pair states (PEPS)\cite{peps1,peps2} network,
where the tensors living on each physical site are coupled by nearest-neighbor links. 
These higher dimensional tensors split the entanglement uniformly over many bonds, 
so the network satisfies the area law regardless of the width of the system. 
On the other hand, the presence of loops in the network, and the higher number of indices on
each tensor lead to much higher complexity in terms of the number of states per link, compared to DMRG. 
Currently, for many 2D problems, DMRG and PEPS are complimentary. 
 
\begin{figure}[t!]
\includegraphics[width=0.48\textwidth]{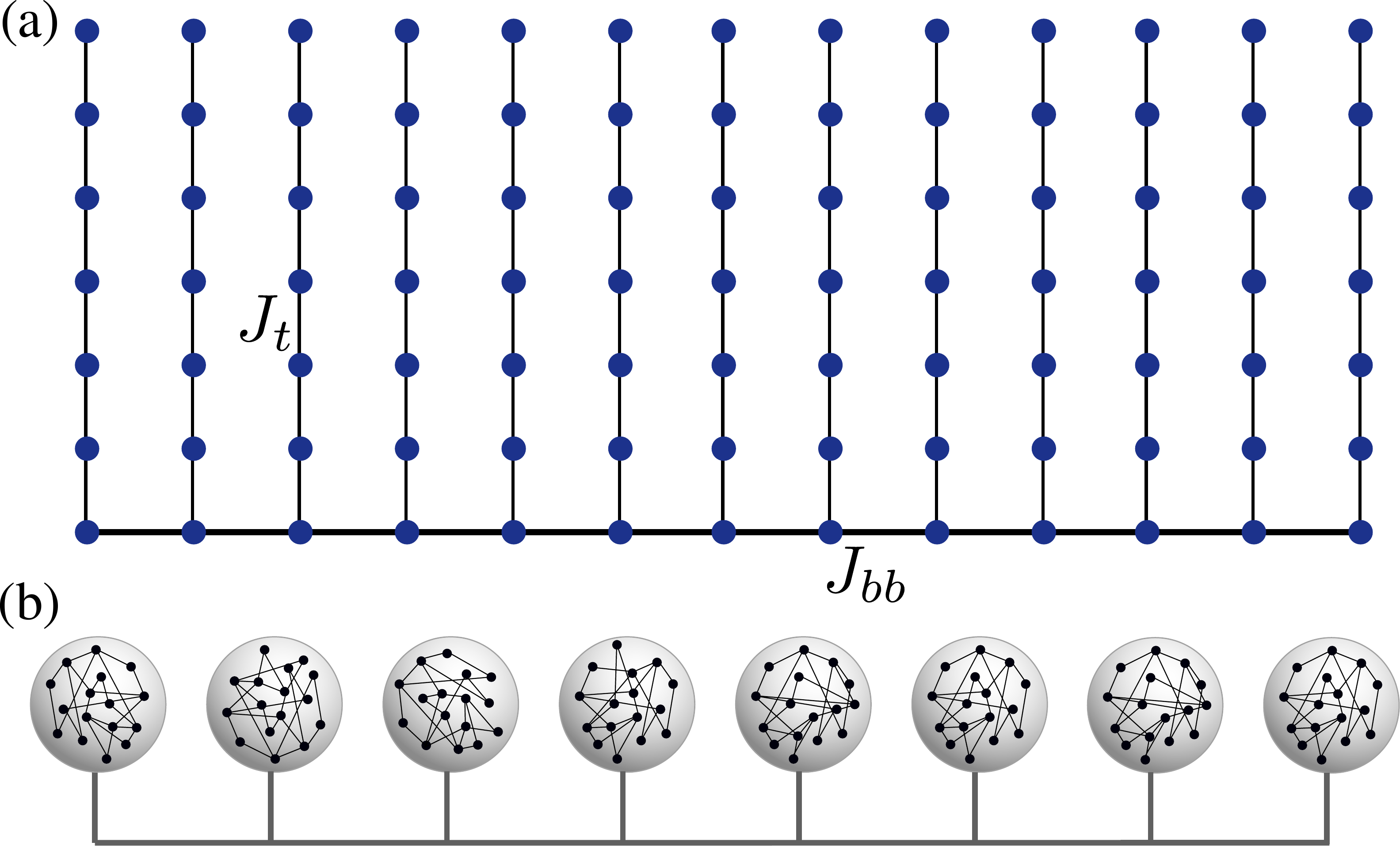}
\caption{(Color online) 
Sketches of the lattice geometries suitable for a comb tensor network. (a) The
simplest comb lattice that essentially repeats the geometry of the tensor
network. Each dot represent a lattice site with local degrees of freedom
and links represent an interaction between nearest neighbors. (b)
A one-dimensional chain of complicated zero-dimensional objects. The
entanglement between the clusters is assumed to satisfy a one-dimensional
area law. } \label{fig:comb_sketch}
\end{figure}

In this paper we study another tensor network
that goes beyond 1D---the comb, shown in Fig.\ \ref{fig:comb_sketch}.
A comb consistes of a 1D backbone, with 1D teeth projecting out from it.
A comb is a special case of a tree tensor network; it can also be considered
a generalization of a Y-junction network, for which DMRG algorithms were developed for Heisenberg
spin systems \cite{PhysRevB.74.060401}. It is also related to fork tensor networks introduced recently for
multi-orbital Anderson impurity models \cite{PhysRevB.81.125126,forktn}.  
It may be that the physical interactions and sites in a system appear in a comb geometry, in which
it is natural to use a comb tensor network to study the system---see Fig.\ \ref{fig:comb_sketch}(a). 
Another, less obvious case where a comb may be particularly useful is  for a 1D system composed
of complicated, highly entangled but finite units.  In this case a comb geometry can be effective
in isolating the intra-unit entanglement into a tooth, leaving the backbone less entangled---see
Fig.\ \ref{fig:comb_sketch}(b). To be studied with the
comb tensor network the model has to satisfy the one-dimensional area law (at
least along the backbone); otherwise the algorithm has the same limitations as a
snake-like DMRG for a two-dimensional system.

The motivation to develop a comb tensor network is two-fold. From a numerical
point of view, the complexity of the algorithm scales with an auxiliary bond
dimension $D$ as $D^5$, which is much lower than two-dimensional PEPS. On the
other hand, for a fixed bond dimension the complexity of the comb tensor
network is higher than in DMRG. However, the higher dimensional tensors
placed along the backbone, split two channels of entanglement: one within the
clusters and one along the backbone. In DMRG the entanglement from these
two channels is superimposed, which results in a large bond dimension. Thus,
when the entanglement along the backbone and within the teeth are of the same
order of magnitude, the comb tensor network requires much lower bond dimension
than conventional DMRG. 

From the point of view of possible applications, the comb tensor network is also
very promising. First of all, a comb geometry itself can 
lead to exotic phases and phase transitions even for the simplest
Heisenberg model. Among them are various junctions of spin chains
\cite{PhysRevB.74.060401,2018arXiv181106794B} and quantum wires
\cite{PhysRevB.79.085122,critical_teeth} that have attracted a lot of attention over
the past decade.  Modern technologies allow one to realize comb lattices in 
cold atom and adatom experiments.
In addition, combs may be used to study coupling effects in 
clustered materials, such as the recently discovered 24-spin boleite\cite{boleite}.

The comb tensor network may also be use for quantum chemistry. There are molecules which
have backbone and chain geometries resembling combs, such as 
triglycerides or polysulfones\cite{polysulfones}, for which one might
make Hubbard-like models.
At the more ab initio level, consider quantum chemistry DMRG, where the sites in the 
algorithm are orbitals, including core orbitals as well as valence. The entanglement within
a core gets added to the interatomic entanglement. It would be natural
to have the core degrees of freedom of an atom appear as sites of a tooth, with the
atoms connected along the backbone, separating the two types of entanglement.
In a similar spirit, associating the teeth of the comb with finite-size
Sachdev-Ye-Kitaev (SYK)\cite{sachdev_ye,kitaev,sachdev_ads_cft,mcgreevy} clusters,
intercluster entanglement can be dealt with separately.

The rest of the paper is organized as follows. In Sec.\ \ref{sec:state_as_a_comb}
we will set up the basic properties of the comb tensor network, including the
transformation of the wave-function between the comb and the traditional
matrix-product-state forms and the mixed canonical form of the comb network. In 
Sec.\ \ref{sec:varopt} we provide further details on the numerical approach
and discuss the variational optimization algorithm of the wave-function as a
comb tensor network. In Sec.\ \ref{sec:spin1} we apply these algorithms to
study the low-energy properties of the Heisenberg spin-1 comb lattice. We
summarize our results in Sec.\ \ref{sec:conc}.

\section{Quantum state as a comb network}
\label{sec:state_as_a_comb}

\subsection{Mixed canonical form}

Putting an MPS into a {\it mixed canonical form} (MCF) simplifies drastically both
optimization and the following re-usage of the wave-function\cite{dmrg4}. The MCF is characterized
by the combination of left-orthogonal tensors on the left and 
right-orthogonal tensors on the right, with a site or link between them called the orthogonality center (OC). 
If the OC is a link, an extra two-index diagonal tensor is associated with the OC.
No orthogonality is associated with OC itself.
In variational optimization, the MCF simplifies a generalized diagonalization to a 
simple one,  and fixes the norm of the wave-function.
When extracting local observables, the MCF allows one to skip the
contraction of the complete tensor network, utilizing only those tensors that
are located at and between the local operators of interest\cite{dmrg3}.

\begin{figure}[t!]
\includegraphics[width=0.49\textwidth]{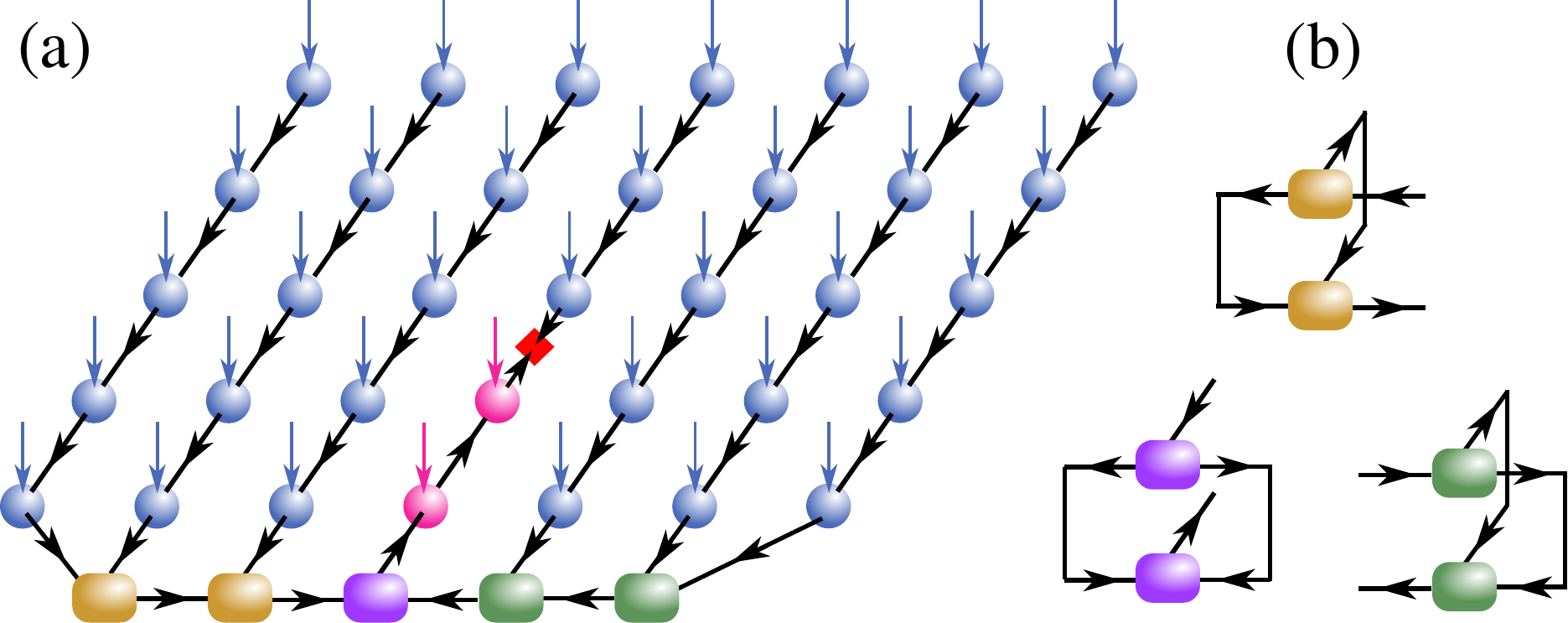}
\caption{(Color online) (a) Mixed canonical form of a comb tensor network.  Circles are
 MPS tensors, rectangles are auxiliary backbone tensors. A red diamond
 indicates a diagonal link OC matrix. Arrows indicate direction of orthogonality
and also how to keep track of Abelian symmetries, as explained in the text.
(b) Graphical representation of various normalization condition for auxiliary tensors on a backbone. Each pair of tensors contracted over connected lines corresponds to the identity matrix of dimension equal to the bond dimension of non-contracted legs. 
} \label{fig:canonical}
\end{figure}

In analogy with a one-dimensional MPS we introduce an MCF for the comb tensor network. 
To reduce the complexity of the algorithm we associate any physical degrees of freedom
on the backbone to the first sites of the teeth, so that the backbone tensors, as well as
the tooth tensors, only have three indices.
Let us consider the case where the OC is on a tooth, and specifically on a link, as shown 
in Fig.\ \ref{fig:canonical}. 
The notion of the left and right normalized tensors of an MPS is generalized for the comb.
We attach arrows to each link, with each arrow pointing along the path connecting to the OC. 
For each tensor, one link (the forward link) points away from the tensor.
To give the orthogonality condition, we contract a tensor with its conjugate, 
contracting over all indices except the forward link. The resulting two index tensor is equated
to the identity tensor, as shown in Fig.\ \ref{fig:canonical}. 
Note that a Hermitian conjugated tensor has its arrows reversed.

In Fig.\ \ref{fig:canonical}(b) we show graphical representations of the three types of
orthogonality conditions.
As in the case of MPS, the MCF simplifies the diagonalization in the variational optimization at each step from
a generalized to a simple eigenvalue problem, reducing the
computational cost and stabilizing the algorithm.  
The MCF also simplifies measurement of local
observables; for an operator on a tooth, the complexity is equivalent to the standard DMRG
complexity $D^3$. Computing correlation functions along the backbone has higher 
complexity $D^4$, however the contraction of the network on each tooth can
be skipped and the complexity does not depend explicitly on the length
of the teeth.
%

%

%
 

\subsection{From an MPS to a comb}

\begin{figure}[t!]
\includegraphics[width=0.4\textwidth]{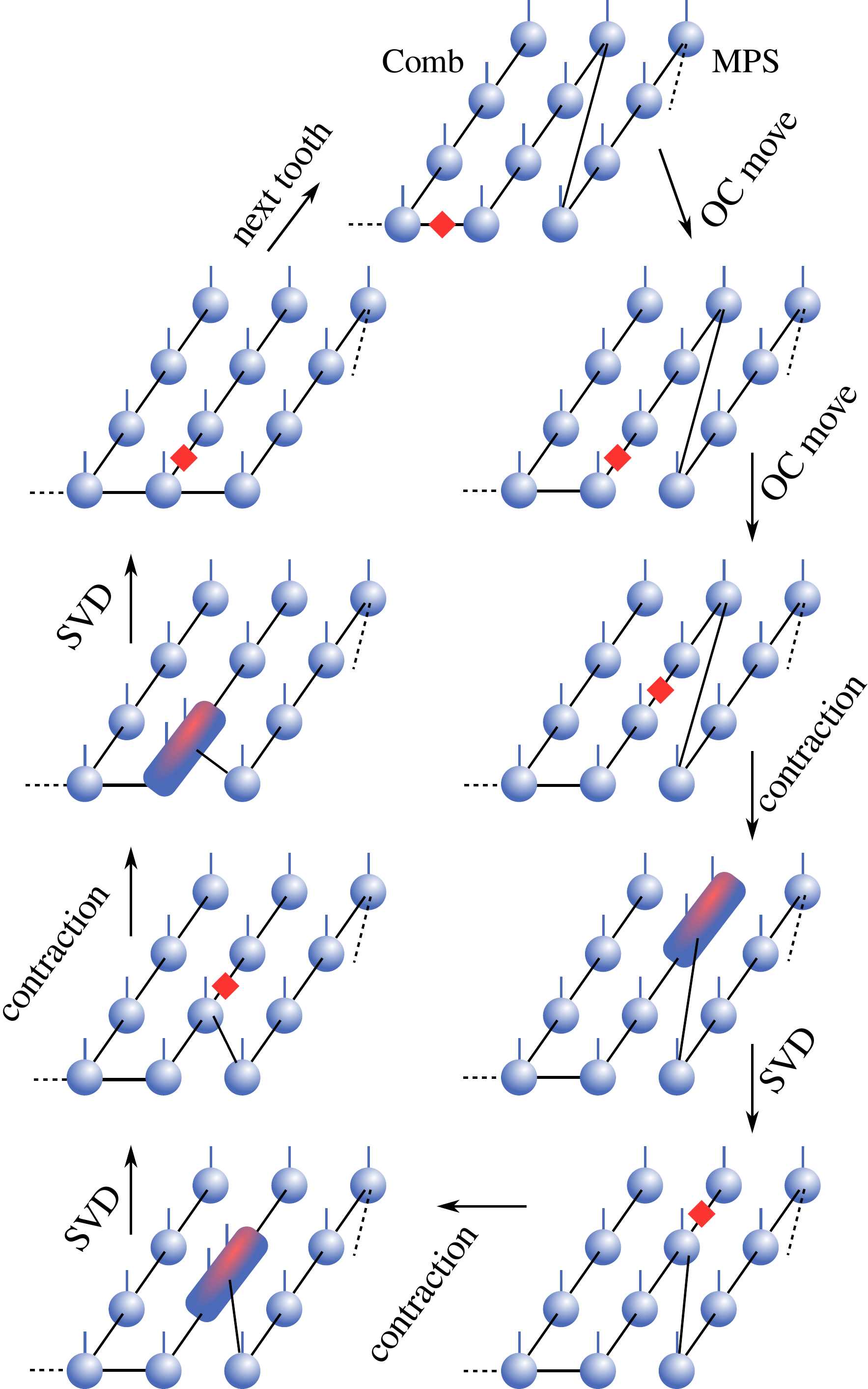}
\caption{(Color online) Example of the MPS to comb tensor network transformation
via contraction and SVD decomposition. We show a complete transformation
of one tooth, which can be repeated tooth by tooth along the backbone. 
The orthogonality center (OC) must be moved to the tooth being transformed,
so that any pair of tensors being contracted include the OC. This assures that the
singular values obtained after the SVD decomposition are equivalent to the
Schmidt values and thus allow optimal truncation.} \label{fig:mps_to_comb}
\end{figure}

In principle, an arbitrary tensor network representation of a wavefunction on a finite number of sites 
can be transformed into an MPS, and vice-versa. The resulting MPS may, however, have large bond
dimension $D$, an example being the conversion of a PEPS wavefunction for a 2D cluster into an MPS. 
Even in cases where the bond
dimension of both represenations is modest, it may not be so easy to transform back and forth,
particularly if the tensor network has loops. An advantage of the comb network is that there
is a very simple, robust algorithm for transforming back and forth between a comb and an MPS,
using simple contractions and SVDs.
In Fig.\ \ref{fig:mps_to_comb} we illustrate this transformation.
Note that the orthogonality center
(OC) has to be moved to any sites being contracted, so the SVD decomposition corresponds to
the Schmidt decomposition of the wave-function. 
Although the figure shows the MPS as living on a 2D square lattice, only the 1D connections of the MPS
are important for this transformation algorithm. 
As immediately follows from
Fig.\ \ref{fig:mps_to_comb}, the complexity for this transformation scales with
the bond dimension as $D^4$, where we have not distinguished between bond dimensions along
the backbone versus a tooth.

\section{Variational optimization}
\label{sec:varopt}

The direct mapping between the comb tensor network and
an MPS described in the previous section allows one use standard DMRG on the MPS
to get a comb ground state, but in general it is more natural and probably more efficient
to optimize within the comb geometry.
%
Since the comb has no loops, the optimization is similar to that of DMRG,
but there are specific techniques for the teeth versus the backbone sites.
In addition, it is useful to represent the Hamiltonian in a comb tensor network
form.  Not wishing to introduce an additional term for this comb operator,
we will call this a projected entangled pair operator (PEPO), borrowing
from projected entangled pair state (PEPS) terminology.
There are five main ingredients in the
proposed algorithm: the representation of the Hamiltonian as a PEPO,
the creation of a good initial
wave-function, the recipe for a full sweep to update all tensors in the network, 
the recipe for a back-bone sweep
to update only auxiliary tensors along the backbone, and efficient
measurements of observables.

\subsection{Hamiltonian}
Let us consider as an example an $N\times L$ comb lattice shown in
Fig.\ \ref{fig:comb_sketch}(a) with Heisenberg nearest-neighbor interaction with
coupling constant $J_t$ along the tooth and $J_{bb}$ along the backbone:
\begin{equation}
  H=J_{bb}\sum_{i=1}^{N-1}{\bf S}_{i,1}\cdot{\bf S}_{i+1,1}+
  J_t\sum_{i=1}^N\sum_{j=1}^{L-1}{\bf S}_{i,j}\cdot{\bf S}_{i,j+1},
  \label{eq:comb_hamilt}
\end{equation}
where $i$ indicates which tooth and $j$ indicates the site 
within the tooth, starting from the backbone. The construction of
the MPO for $2\leq j\leq L$ is standard and is well described, for example, in
Ref. \onlinecite{dmrg4}. For completeness we provide the explicit form of the MPO. For
$2\leq j\leq L-1$ it is given by
\begin{equation}
H(i,j)=\left( \begin{array}{ccccc}
I & . & .& . & . \\
S^- & . & .& . & . \\
S^+ & . &  . & .& . \\
S^z & . & .&  .& . \\
hS^z & J_tS^+ & J_tS^- & J_tS^z  & I \\
\end{array} \right),
\end{equation}
where $I$ is an identity matrix, $h$ is an external magnetic field, and
zero-elements are marked by dots for clarity. Both $I$ and $S$ are operators;
if the operators are written in terms of indices then
the total rank of the tensor is four. At the tip of the tooth $j=L$ the
tensor is given by the first column of this matrix and has rank three.  At the
backbone the rank of the on-site tensors that represent the Hamiltonian
is higher, because on top of two physical bonds, they contain three auxiliary
bonds that connect them to all their neighbors. 
Since there are no loops in the comb, the construction of
these PEPO operators is as simple as the conventional construction of the MPO.
In the tensor $H_{\alpha\beta\gamma}$ we use the label $\alpha$ for the left,
$\beta$ for the upper and $\gamma$ for the right auxiliary legs. 
The tensor is very sparse, but we can represent the slice $\beta=1$ as a matrix,
and then the only nonzero elements left have $\alpha=5$ and $\gamma=1$, so
these values can be written as a vector in the index $\beta$:
\begin{multline}
H_{\beta=1}=\left( \begin{array}{ccccc}
I & . & .& . & . \\
S^- & . & .& . & . \\
S^+ & . &  . & .& . \\
S^z & . & .&  .& . \\
hS^z & J_{bb}S^+ & J_{bb}S^- & J_{bb}S^z  & I \\
\end{array} \right), \\
H_{5,\beta,1}=[hS^z,J_tS^+,J_tS^-,J_tS^z,I] \ \ \ \  \ \ \ \  \ \ \ \  \ \ \ \  \ \ \ \  \ \ \ \\
\label{eq:PEPO}
\end{multline}

\begin{figure}[t!]
\includegraphics[width=0.2\textwidth]{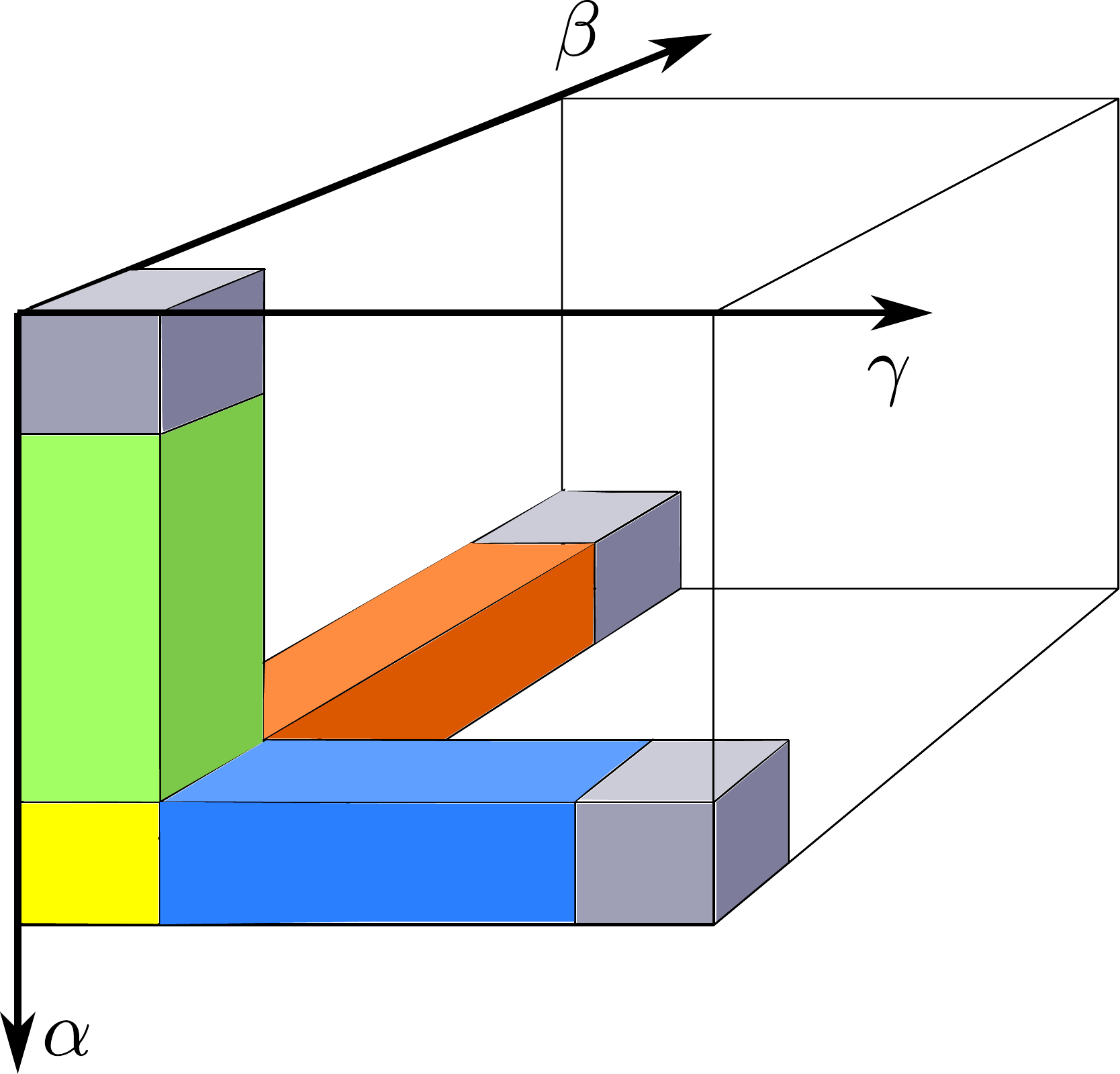}
\caption{(Color online) Pictorial representation of non-zero elements of the
PEPO tensors on the backbone. Grey boxes label the identity matrix $I$,
yellow boxes label the external field operator $hS^z$, blue box correspond
to three elements $[J_{bb}S^+, J_{bb}S^-, J_{bb}S^z]$, green boxes contain
the elements $[S^-,S^+,S^z]$, and the orange box corresponds to $[J_tS^+,
J_tS^-, J_tS^z]$. Empty space corresponds to zero elements.}
\label{fig:PEPO}
\end{figure}

We find it instructive to show the pictorial representation of non-zero
elements of the backbone PEPO tensor (see Fig.\ \ref{fig:PEPO}), where the gray boxes state for
identity operators, the green box includes three elements of the first column of
the matrix in Eq. \ref{eq:PEPO} $[S^-, S^+,S^z]$, the blue box contain three elements
of the last row of this matrix $[J_{bb}S^+, J_{bb}S^-, J_{bb}S^z ]$, the orange
box encodes three elements for the interaction within the tooth $[J_bS^+,
J_bS^-, J_bS^z ]$, and the yellow box corresponds to the field-term, which is set to
zero throughout a paper. Of course, this form of the Hamiltonian is
model-dependent, and if the interaction is more complicated and extends beyond
nearest-neighbors the structure of the PEPO is not as simple.

\subsection{Sweep}

Let us first explain the variational optimization of the comb tensor network
starting from some initial state. Later, we shall come back to the creation of a
good initial wave-function.

\begin{figure}[t!]
\includegraphics[width=0.48\textwidth]{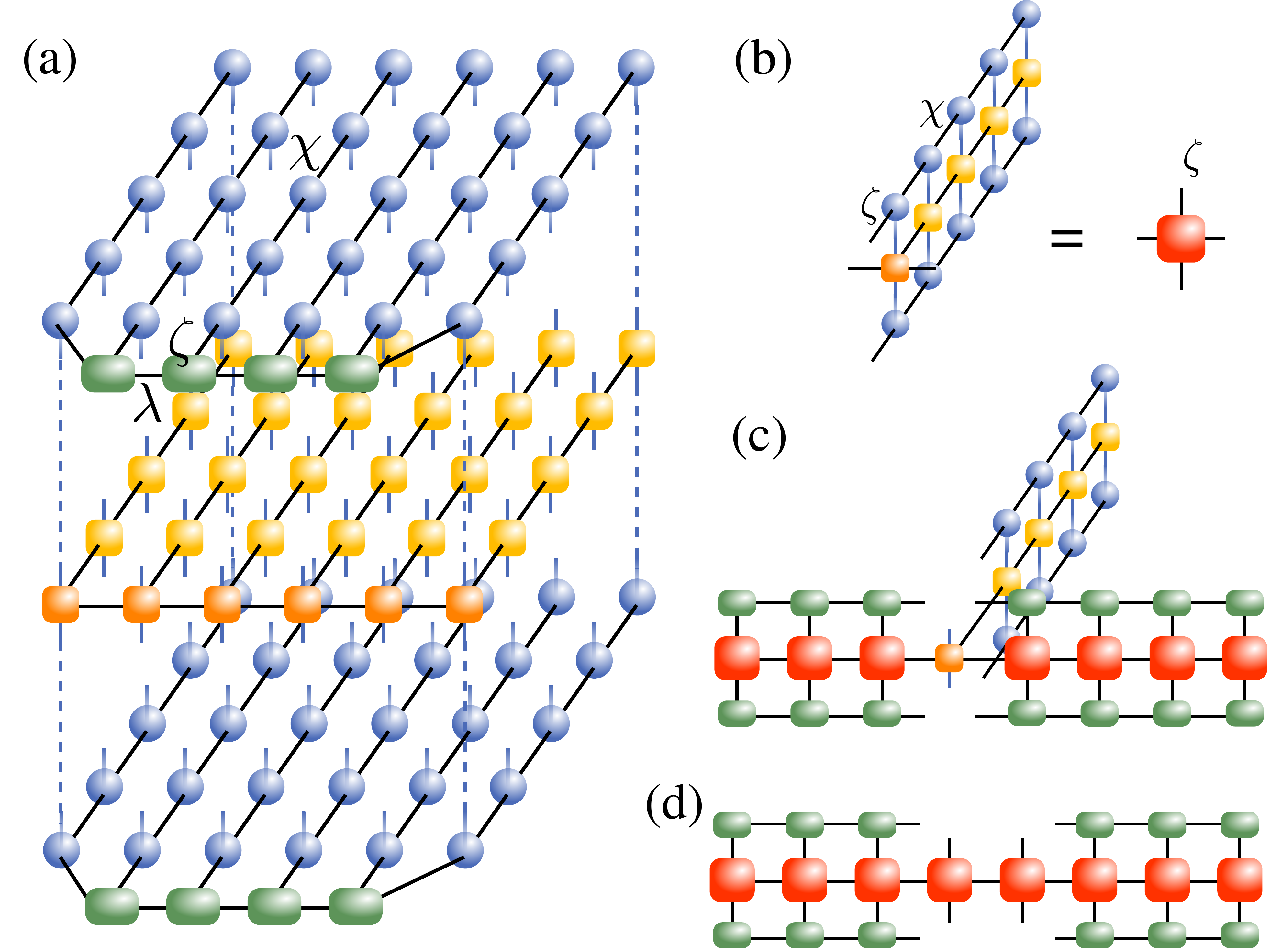}
\caption{(Color online) (a) Full tensor network that evaluates the total energy. 
The top and bottom layers are the wave-function and its conjugate, while the middle
layer is the Hamiltonian. Vertical indices are physical degrees of freedom.
Note that the structure of the Hamiltonian PEPO is different from the comb wave-function
tensor network, since the wave-function puts the backbone connections on auxiliary tensors (green boxes). (b) Fully contracted tensor network on a single tooth can be interpreted as an effective PEPO with extended bond dimension. (c) Connect update: simultaneous update of auxiliary backbone tensor and first MPS tensor on a tooth. (d) Backbone update: simultaneous update of two auxiliary tensors. }
\label{fig:optimization}
\end{figure}

Fig.\ \ref{fig:optimization}(a) shows the full
tensor network that evaluates the total energy consisting of a ket wave-function $|\psi\rangle$ (lower
surface of blue and green tensors), the Hamiltonian in local representation
(middle surface with yellow (MPO) and orange (PEPO) tensors)  and a bra vector
$\langle\psi|$, represented by the upper surface of tensors. The contraction over
all physical and auxiliary links gives the total energy
$E=\langle \psi|H|\psi\rangle$. We assume that the wave-function is always
written in a canonical form defined above, so the normalization can be omitted
here. Note that the number of tensors that represent the wave-function is larger
than the number of tensors that represent the Hamiltonian, due to the use of
auxiliary backbone tensors without physical indices. 
So,  PEPO backbone Hamiltonian tensors are contracted with the first MPS tooth tensors.

In a full sweep we go through the whole network shown in Fig.\ \ref{fig:optimization}(a) and update the on-site tensors iteratively. During the first half-sweep we move from left to right, and then reverse during the second half-sweep. Within each tooth we move from the backbone to the edge and back. At each iteration we update two on-site tensors. A generalization to single- or multiple-site update is straightforward. 

In most of the cases the update can be reduced to the diagonalization of the effective Hamiltonian that consists of the left and right environments and a pair of MPO between them. So on a tooth this is standard 1D DMRG. For a pair of auxiliary backbone tensors, one can view a fully contracted tensor network on a tooth as an effective MPO as shown in Fig.\ \ref{fig:optimization}(b). The effective Hamiltonian for two auxiliary tensors also looks like the form in 1D DMRG; see Fig.\ \ref{fig:optimization}(d).
 However, there is a special case that cannot be reduced to the standard network with two environments and two MPOs. When an auxiliary tensor is updated together with the first MPS tensor on a tooth (a connect update) the effective Hamiltonina contains three environment each of which is connected to a PEPO in the middle, as shown in Fig.\ \ref{fig:optimization}(c).

Depending on the model, we find it useful to alternate the full sweep with a 
simplified one that updates only the backbone tensors. In particular this make
sense when the backbone chain is critical while the teeth are gapped. This short
sweep consists only of backbone updates.

We increase the number of states every half sweep. We control three parameters
that restrict the number of states: $\chi$ is the bond dimension within the tooth;
$\zeta$ restricts the bond in connect updates; and $\lambda$ controls
the dimension of a bond along the backbone.  When all bonds are
equivalent $\chi\approx\lambda\approx\zeta\approx D$ the complexity of the
backbone update is $D^5$, the complexity of the connect update is $D^4$, and the
complexity of the tooth update is only $D^3$, as in the standard DMRG. Of course, the
complexity is very different when the number of states on the tooth and on the
backbone are of a different order of magnitude. For example, if the state on the
tooth is close to the Affleck-Kennedy-Lieb-Tasaki state\cite{AKLT}, the bond dimensions
$\chi$ and $\zeta$ are small in comparison to $\lambda$ and the complexity is
only $\lambda^3$.

\subsection{Initial wave-function}

We start our simulation by forming a rough initial wave-function. As in 1D DMRG there are various different approaches and depending on the problem one or another can be preferable. 
The simplest one is to start with a product state - tensor network with auxiliary bond dimension $D=1$. If Abelian symmetry is used, the chosen product state should satisfy it. This approach is efficient assuming one slowly increases the bond dimension in multiple sweeps.

In variational optimization of an MPS it is well established that one of the most accurate and unbiased guesses can be obtained through infinite-size DMRG\cite{dmrg4}. It starts with small clusters that can be solved exactly, say four spins,  and at each iteration the size of the 1D chain increases by two spins. Infinite-size DMRG produces a guess wave-function with finite bond dimension (typically $D\approx 10-50$) that naturally preserves the symmetry of the network with respect to the center of a chain and provide an excellent starting point for the following finite-size routine. 
We find that an excellent initial wave-function comes from using infinite-size DMRG on a chain with $2L$ sites to produce an initial guess for the MPS tensors on the teeth. The Hamiltonian used in infinite-size DMRG should be as close as possible to the original MPO on a tooth. 
One can translate the $2L$ MPS tensors to define the tensors of two adjacent teeth. Since this wave-function ignores the backbone interactions, one can slightly truncate  the MPS bond dimension at the backbone (the middle) to make a subsequent initial set of backbone sweeps fast and inexpensive. These backbone sweeps can start either with the product-teeth states, or with an infinite-size DMRG along the backbone. 

One can also use an infinite-size DMRG to initialize the backbone after constructing guesses fir the tooth tensors. By treating a fully contracted tooth as an MPO with large 'physical' bond
dimension, one can initialize the auxiliary guess tensors by performing a standard infinite-size DMRG on the backbone.

In the present paper we produce a guess wave-function by performing infinite-size DMRG in both directions.

\subsection{Implementation of Abelian symmetries}

When allowed by the model, the implementation of Abelian symmetries in the
comb tensor network allows one to select a specific symmetry sector and thus to
compute the energy gap to magnetic excitations in a trivial way. Moreover, this reduces significantly an effective size of the
Hilbert space and therefore speeds-up the convergence at each iteration.
Finally, in the presence of a U(1) symmetry the tensors have block-diagonal structure that simplifies their storage and access. 

In order to keep track of Abelian symmetry, one can group the states on each index by quantum number (say $S^z$).
Then one stores only nonzero blocks in the tensors, and
the arrows in the diagrams in Fig.\ \ref{fig:canonical}(a) indicate how one adds up the quantum numbers. For the nonzero blocks, 
the sum of the in-going quantum numbers must equal
the outgoing quantum number on each tensor. At the OC, where all arrows point inward, one can regard
the OC tensor as a wave-function in an orthonormal basis.  Each link connecting to the OC tensor
represents an orthonormal set of basis states living on the associated branch of the comb, 
and the whole basis is the direct product over all the branches. 
The total quantum number of the OC-tensor wavefunction is the sum of all
incoming quantum numbers. If OC is associated with the tensor that contains physical index, its contribution should also be added to the total quantum number. All nonzero blocks of the OC tensor have this same quantum number.


\subsection{DMRG versus comb}

Let us now compare the complexity of optimizing the comb versus an equivalent MPS.
First, we will consider a square spin-$1/2$ comb with antiferromagnetic nearest-neighbor Heisenberg 
interactions,  where we set both coupling constants of
the Hamiltonian (\ref{eq:comb_hamilt}) to be equal $J_{bb}=J_{t}=1$.

\begin{figure}[t!]
\includegraphics[width=0.49\textwidth]{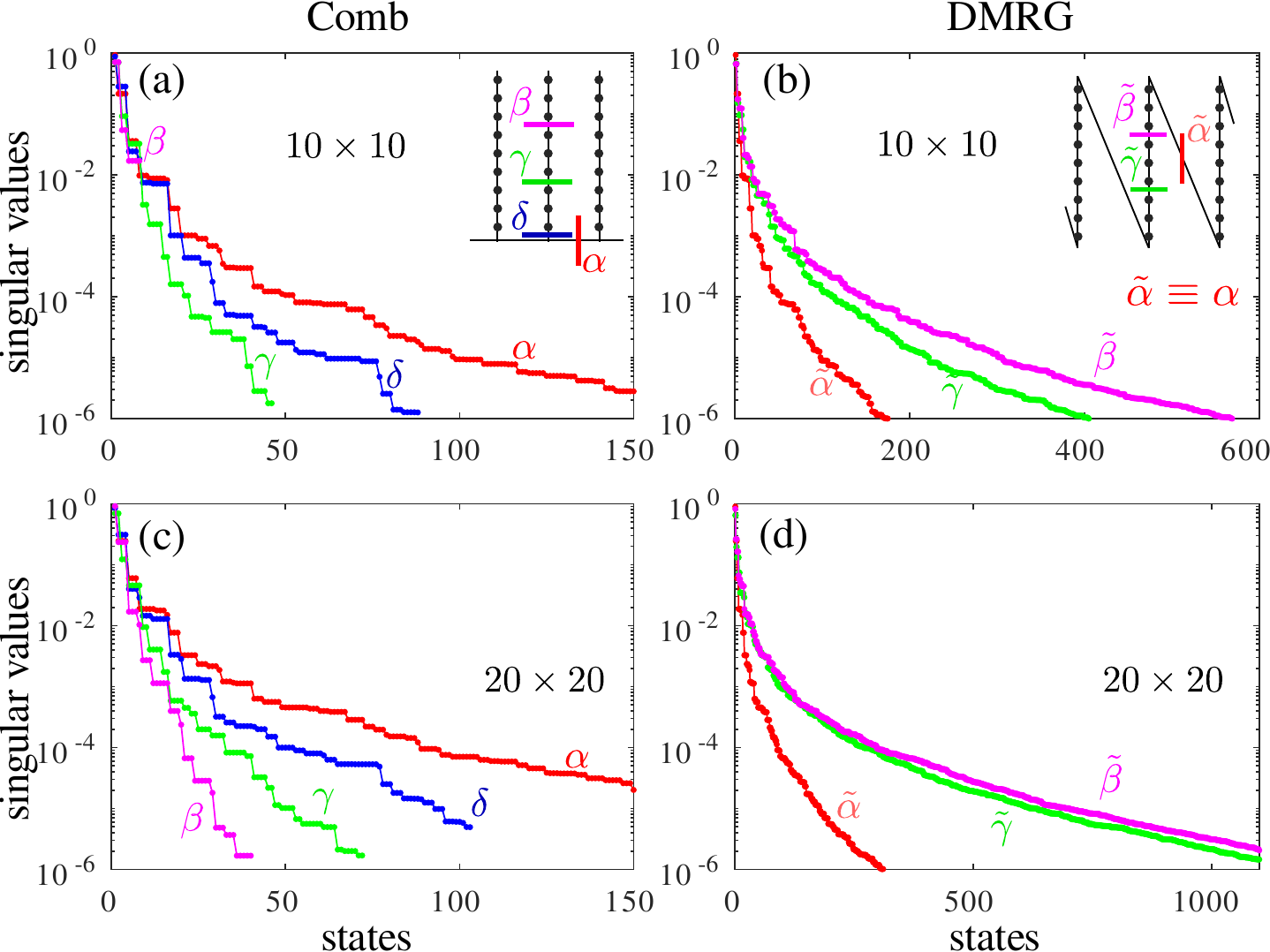}
\caption{(Color online) Singular values for square spin-1/2 comb with (a-b)
$10\times10$  and (c-d) $20\times20$ sites for various bi-partition of the
system distinct for the comb tensor network (a,c) and for the DMRG (b,d).
Red - cut across the backbone in the middle of the comb, blue - cut of the
whole tooth in the middle on the comb from the rest of the network, green - cut between (a-b) sites 3 and 4 and (c-d) sites 7 and 8 of the middle
tooth, magenta - cut between sites (a-b) sites 7 and 8 and (c-d) sites 14 and
15. The insets show the schematic position of the corresponding cuts. Note the difference in the $x$-scale in left and right panels; red lines are the same on left and right panels and can be taken as a reference. }
\label{fig:singular_values}
\end{figure}

The algorithms split the system into two parts in different ways. It is useful to
compare the singular values (square root of the Schmidt values) for these various cuts.
Results for $10\times 10$ and $20\times 20$ combs
are shown in Fig.\ \ref{fig:singular_values}. Note that in both algorithms
the cut $\alpha$ ($\tilde\alpha$) across the backbone leads to the same
bi-partition of a lattice and thus the singular values are equal (red dots). In
variational optimization of a comb tensor network it is natural to cut all or part of a tooth 
from the rest of the system as sketched in the inset of
Fig.\ \ref{fig:singular_values}(a). Here we show singular values for three
example cuts: the whole tooth, which is the same as tooth-connect (cut $\delta$, blue
color); a two-thirds cut ($\gamma$, green color); and a one third cut ($\beta$,
magenta color) of the tooth. We see from
\ref{fig:singular_values}(a,c) that the entanglement is
largest along the backbone and, as expected, decays very fast upon approaching the end of the
tooth.

By contrast, in DMRG applied to a comb lattice the entanglement is smallest on the bond that connects the
upper and lower edges of two teeth. It is significantly larger within the
teeth, where the entanglement from both the backbone and from within the tooth are present.
We stress that the drastic difference between the
decay of Schmidt values in DMRG and the comb network is caused by the different cuts,
not any details of the optimization.

\begin{figure}[t!]
\includegraphics[width=0.49\textwidth]{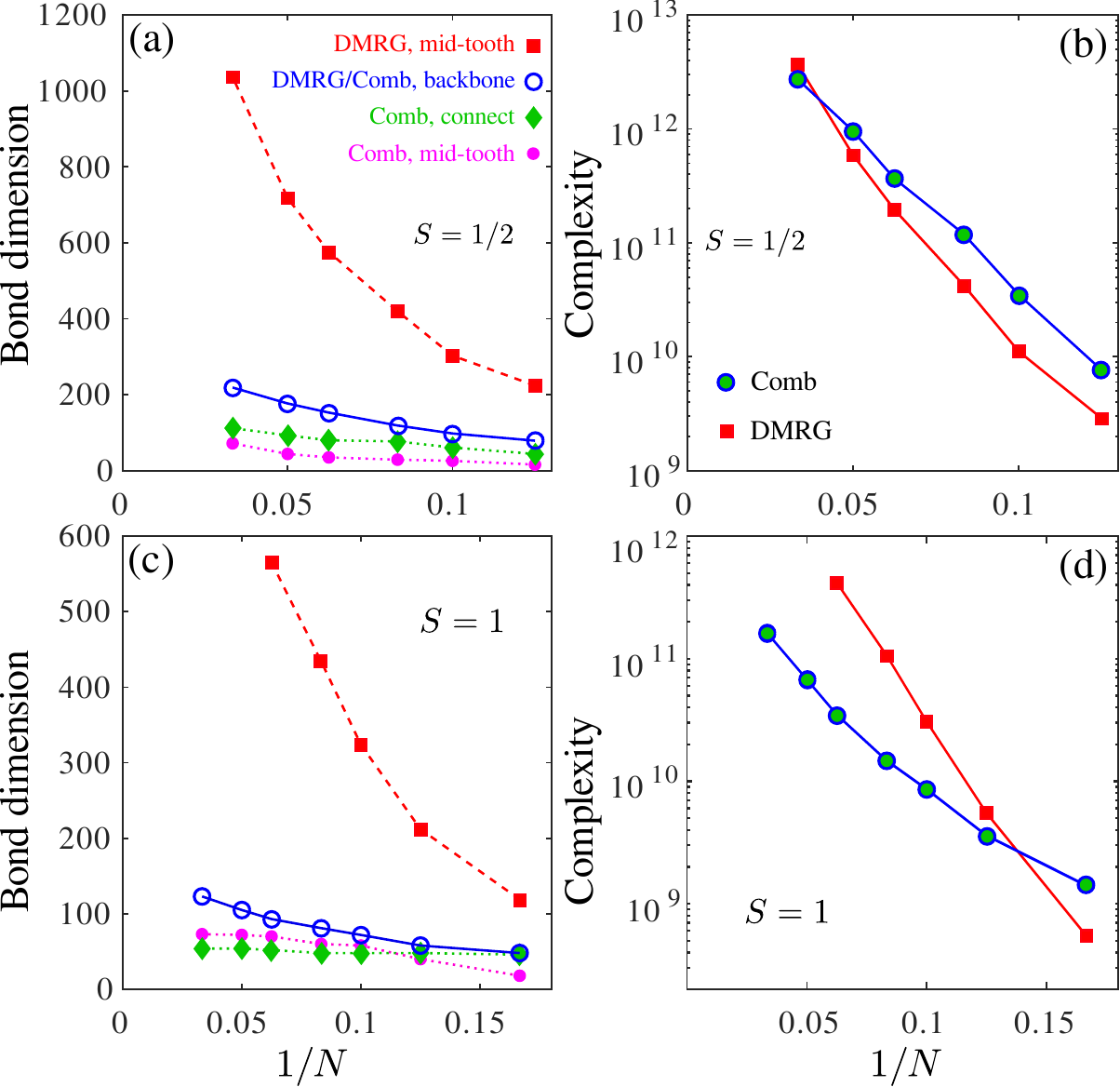}
\caption{(Color online) (a) Bond dimensions as a function of linear size of a
square $N\times N$ spin-1/2 comb lattice with Heisenberg nearest-neighbor
interaction for various cuts natural for either DMRG, the comb,  or both
optimization schemes. A cut across a backbone has the same bond dimension in
both cases and is marked with blue circles; red squares are for
the cut across the middle of a tooth which is natural for
DMRG; green diamonds are for a cut between the whole tooth and
the backbone; and magenta circles are for a cut in the middle of a
tooth in the comb network. (b) Estimate of the leading term in the
complexity of DMRG and variational optimization of a comb based on the data
from panel (a). (c-d) same as (a-b) but for spin-1 comb lattice with
$J_t=1$ and $J_{bb}=0.1$, and where the end spin of each tooth is replaced by a
spin-1/2 to lift the macroscopic quasi-degeneracy of the ground-state due
to edge states.  }
\label{fig:bond_dimension}
\end{figure}

We measure the number of states necessary at each bond to  keep the truncation
error below $10^{-4}$ for singular values (and $10^{-8}$ for Schmidt values).
The results are summarized in Fig.\ \ref{fig:bond_dimension}(a). The bond
dimension along the backbone $\lambda$ grows slowly with the system size. Such
slow decay qualitatively agrees with the slow logarithmic divergence of the
bond dimension for critical systems. The same is true for the bond within the tooth
and the bond that connects the tooth with the backbone for the comb network. On
the other hand, the bond dimension of the MPS wave-function optimized with DMRG
grows much faster with the system size.

We estimate the complexity of the DMRG and variational optimization of a comb
tensor network based on the bond dimensions. For DMRG the most consuming term
is the contraction of the network in the middle of the tooth with the
single-iteration complexity $D^3$, that has to be performed about $N^2$ times
per half-sweep, so the total complexity is $D^3 N^2$. For a comb tensor network,
the leading term of the complexity is given by simultaneous optimization of two
backbone tensors that has single-iteration complexity $\lambda^3\zeta^2$ and
has to be performed only $N$ times, so the total complexity is given by
$\lambda^3\zeta^2 N$. In the large $N$ limit we expect the comb represenation
to be generally more efficient; for smaller $N$ the comparison depends on the
difference in bond dimensions for the MPS along the backbone versus in the middle of the teeth.

For a spin-1/2 Heisenberg comb, the optimization of the wave-function directly
in a comb geometry gives slightly lower complexity only starting from the
lattice with $30\times 30$ sites as shown in Fig.\ \ref{fig:bond_dimension}(b).
However the two curves follow each other very closely and we expect these behavior
to persist even for larger clusters. 

Now we compare the computational cost of a spin-1 comb with strong coupling
along the teeth $J_t=1$ and weak backbone interaction $J_{bb}=0.1$. The main
difference from the previous case, is that now the teeth are essentially gaped
sub-systems, each of which corresponds to the Haldane finite-size chain. At the
edge of each tooth spin-1/2 edge states emerge. As soon as a backbone
coupling is non-zero these edge states interact between themselves. While
spin-1/2 edge-states emergent at the backbone interact with the coupling
$J_{bb}$, the effective coupling between the edge states at the end of the
teeth is exponentially suppressed with the length of the tooth. This leads to a
massive degeneracy of the ground state with exponentially small splitting
between many in-gap states. In order to avoid this, we replaced spin-1 degrees
of freedom at the end of each tooth by a spin-1/2.  Below we will provide the
detailed study of the model, but here we only focus on the algorithm
efficiency.

Since the teeth are expected to be in a gapped state, according to the area law
the corresponding bond dimension  within the teeth is expected to approach a
constant in the thermodynamic limit. Indeed one can observe it in the  green and
magenta lines in panel (c) of Fig.\ \ref{fig:bond_dimension}. On the other hand,
spin-1/2 edge states are expected to form a critical chain and so the bond
dimension across a backbone  diverges logarithmically with the number of teeth.
In DMRG the entanglement induced by the critical chain is also seen inside the
teeth, so in DMRG each bond dimension diverges with the length of the comb. For the 
chosen set of parameters, comb tensor network turns out to be more efficient
than standard DMRG starting from very small systems of $8\times8$ sites (see
Fig.\ \ref{fig:bond_dimension}). 

To summarize, there exist classes of models, for which the optimization of the
wave-function directly in the comb geometry has lower computational cost than
standard one-dimensional algorithms. Moreover, in some particular cases, e.g.
dimerized state on the backbone, the comb tensor network is less likely to get
stuck in local minima than the DMRG, in which the dimers are  formed
between sites that are far apart.

In the following section we use this algorithm to study the Heisenberg
model on a spin-1 comb lattice.

\section{Spin-1 Heisenberg model on a comb lattice}
\label{sec:spin1}
\subsection{The model}

The spin-1  Heisenberg  chain has long been known to have a  finite bulk
gap\cite{haldane} and  spin-1/2  edge  states\cite{kennedy,hagiwara}. This is
one of the simplest examples of a topologically non-trivial state realized
in spin systems. Here we couple a set of these spin-1 chains into a comb with Heisenberg nearest-neighbors interactions
defined by Eq. \ref{eq:comb_hamilt}.

In the absence of any backbone interaction the edge states of a tooth couple. If the number of spins per tooth is even they
form a singlet and the first excited state is a
Kennedy triplet\cite{kennedy}. When the number of sites on a tooth is odd, the ground-state
is a triplet, and the first excitation is a singlet. In both cases, the energy
splitting between the ground-state and the in-gap excited state(s) decays exponentially
with the length of the tooth. Introducing a backbone interaction, one couples
the backbone edge-states into a spin-1/2 chain. This chain is
then decorated with the spin-1/2 degrees of freedom at the tips with an effective coupling constant that decays exponentially fast
with tooth length. So, for large $L$  the pendant spins cause a macroscopic degeneracy of the ground
state with $2^N$ states, where $N$ is the number of teeth.  To avoid this massive degeneracy, for
most of our calculations we remove the emergent spin-1/2 degrees of freedom at
the end of each tooth by replacing the last spin-1 site by a spin-1/2. We will come back to the model with all spins-1 and study the
effect of the edge spins at the end of the paper.

\subsection{Critical spin-1/2 chain}

When the backbone coupling is small the emergent spin-1/2 edge states, coupled to each other along the backbone, are only slightly changed and form a spin-1/2 chain. According to conformal field theory (CFT) the
Heisenberg spin-1/2 chain is critical and can be described by the
Wess-Zumino-Witten SU(2)$_1$ critical theory. As a confirmation to that, we
extract the critical exponent $d$ from the decay of the Friedel oscillations.
The latter naturally appears in the finite-size chain, because open edges of a
critical spin-1/2 chain favor dimerization, and therefore fix the boundary
conditions. We define the local dimerization order parameter as an absolute
value of a difference between spin-spin correlations on the neighboring bonds
on a backbone:

\begin{equation}
D_{bb}(i,N)=|\langle{\bf S}_{i,1}\cdot {\bf S}_{i+1,1}\rangle-\langle{\bf S}_{i+1,1}\cdot {\bf S}_{i+2,1}\rangle|,
\label{eq:defdim}
\end{equation}
where $i$ is the tooth index and $N$ is the total number of teeth.
Then according to boundary CFT the dimerization decays away from the boundary as:
\begin{equation}
  D_{bb}(i,N)\propto \frac{1}{\left[N\sin(\pi i /N) \right]^d},
  \label{eq:dimprof}
\end{equation}
where $d=1/2$ is a critical exponent of the WZW SU(2)$_1$.

\begin{figure}[t!]
\includegraphics[width=0.49\textwidth]{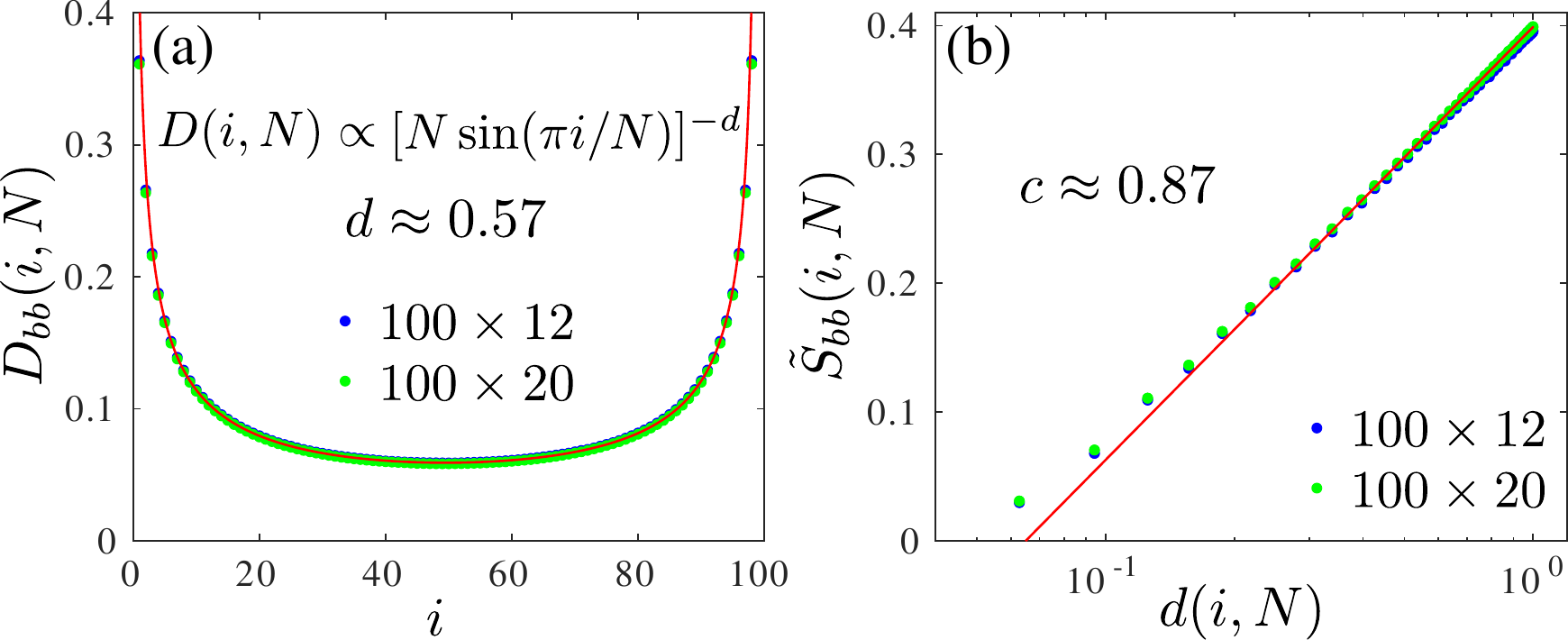}
\caption{(Color online) (a) Dimerization profile along a backbone of a Heisenberg spin-1 comb and  (b) scaling of the entanglement entropy as a function of conformal distance. Results are shown for a comb with 100 teeth and 12 (blue) and 20 (green) sites per tooth, and backbone coupling constant $J_\mathrm{bb}=0.1$. 
}
\label{fig:dimscaling}
\end{figure}

In Fig.\ \ref{fig:bond_dimension} we plot the dimerization profile of the Friedel
oscillations on a comb with 100 teeth and both 12 (blue) and 20 (green) sites
per tooth. One can immediately see that the dimerization profile is independent
of the width of the comb and the data collapse is almost perfect. A fit to the CFT
prediction of Eq. \ref{eq:dimprof} shown in Fig. \ref {fig:dimscaling}(a) gives the critical exponent $d\approx0.54$,
which is higher than the CFT prediction $1/2$ due to the presence of 
logarithmic corrections in  the Heisenberg spin-1/2 chain. 

In addition, we can extract the central charge from the scaling of the entanglement
entropy. Following  Ref. \onlinecite{capponi},  we  define  the  reduced
entanglement entropy on a backbone $\tilde{S}_{bb}(i,N)$  with the Friedel oscillations removed:
\begin{equation}
\tilde{S}_{bb}(i,N)=S_{bb}(i,N)-\zeta\langle {\bf S}_{i,1}\cdot {\bf S}_{i+1,1}\rangle,
\end{equation}
where $\zeta$ is a non-universal parameter. Then,  according  to CFT the
reduced entanglement entropy scales  with  conformal  distance
$d(i)=\frac{2N}{\pi}\sin(\pi i/N)$ as \cite{CalabreseCardy}:
\begin{equation}
  \tilde{S}_{bb}(i,N)=\frac{c}{6}\log d(i)+s_1+\log g
\end{equation}

From the fit shown in Fig. \ref {fig:dimscaling}(b) we find the central charge $c\approx0.87$. This is in reasonable agreement with the $S=1/2$ chain value $c=1$.

\subsection{Crossover}

\begin{figure}[t!]
\includegraphics[width=0.4\textwidth]{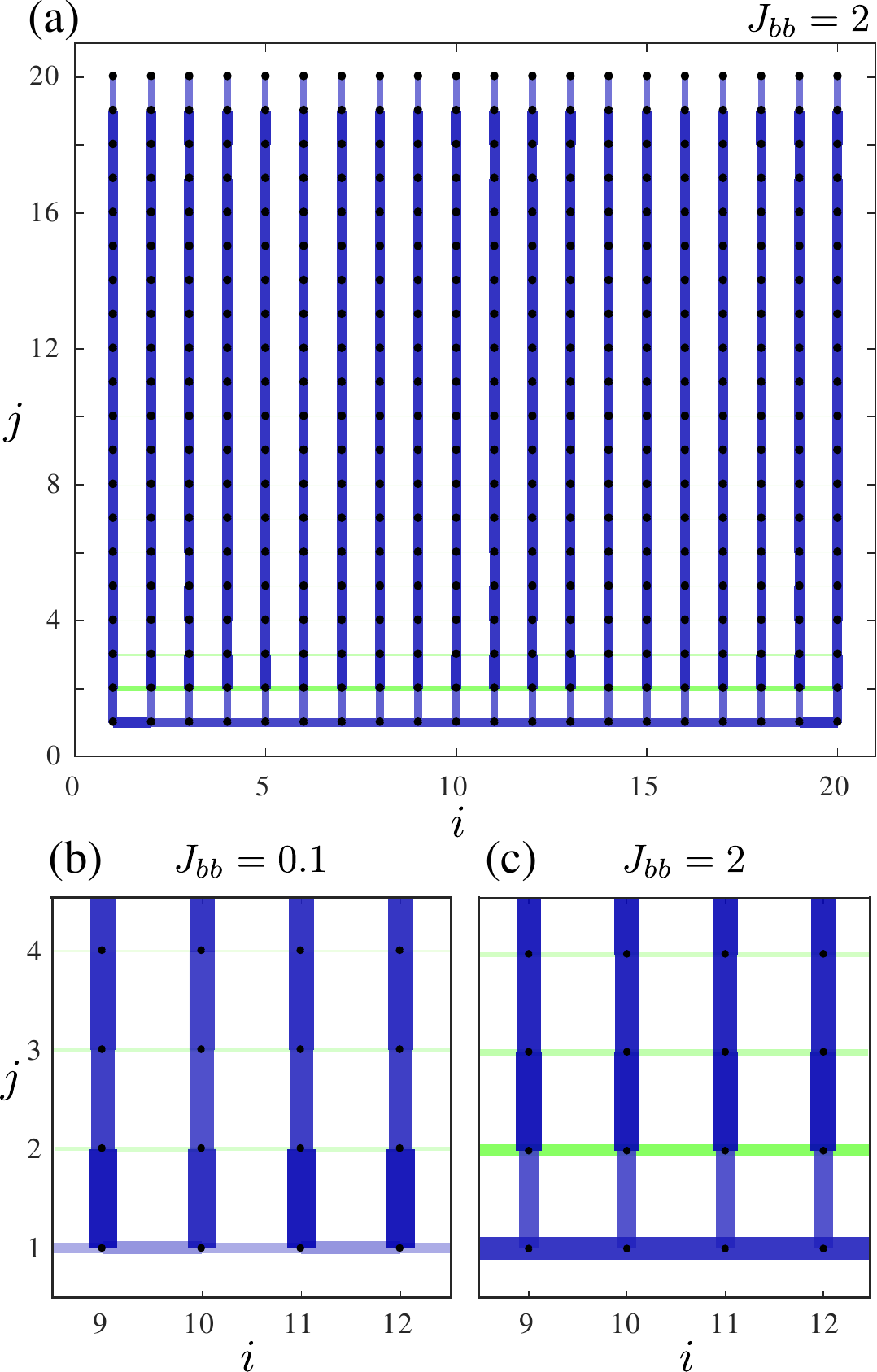}
\caption{(Color online) Correlation graphs that present the strength of the spin-spin correlations $\langle {\bf S}_{i,j}\cdot{\bf S}_{i^\prime,j^\prime} \rangle$ where either $\left[i^\prime=i; \ j^\prime=j+1\right]$ or $\left[i^\prime=i+1; \ j^\prime=j\right]$ in a spin-1 comb. The width and the intensity of the lines are linearly proportional to the strength of the correlations. All correlations have a negative sign. (b) and (c) are enlarged parts of the correlation graphs in the vicinity of the middle of the backbone for (b) $J_{bb}=0.1$ and (c) $J_{bb}=2$, before and after the crossover in $J_{bb}$.}
\label{fig:cor_graph}
\end{figure}

Let us now investigate how the ground-state of a spin-1 Heisenberg comb changes
upon tuning the backbone interaction.  In order to understand how the structure
of the ground-state changes we look at the local correlations between nearest neighbors
on a comb. Fig.\ \ref{fig:cor_graph}(a) shows the strength of the spin-spin
correlations $\langle {\bf S}_{i,j}\cdot{\bf S}_{i^\prime,j^\prime} \rangle$
where either $\left[i^\prime=i; \ j^\prime=j+1\right]$ or $\left[i^\prime=i+1;
\ j^\prime=j\right]$; in other words, we look at the nearest-neighbor
correlation of a square lattice. (The Hamiltonian, however, has couplings only along the comb.) In
Fig.\ \ref{fig:cor_graph}(a) we present our results for a square comb with
$20\times 20$ sites and large backbone interaction $J_{bb}=2$. Note that the
Haldane state is well preserved along the tooth and only a few sites are
affected by the presence of a backbone interactions. We can be even more
specific here, since each tooth has a bulk gap, also in the thermodynamic limit
and the correlation length is finite along the tooth. In this respect, the
presence of the backbone interaction changes the boundary conditions of the
tooth without affecting its bulk properties. So in the next two panels we focus
on the sites in the vicinity of the backbone. In Fig.\ \ref{fig:cor_graph}(b)
and (c)  we show an enlarged part of the correlation graph two values of the
backbone interaction $J_{bb}=0.1$ and $J_{bb}=2$. One can clearly see the
difference between the two: in (b) the strongest correlations are always along
the tooth;  there is a light dimerization along both teeth and the backbone due
to Friedel oscillations induced by open edges; and the correlation between the
teeth above the backbone is negligibly small. In contrast, in (c) the backbone
correlations are almost as strong as the upper part of the teeth, while the
correlation between the first two sites on each tooth are much weaker that
indicates that the backbone is in the Haldane state and is less entangled with
the teeth than in (b). Moreover, the correlation between the second sites of
each tooth is significant, which corresponds to the critical spin-1/2 chain
that is now formed above the backbone.

\begin{figure}[t!]
\includegraphics[width=0.49\textwidth]{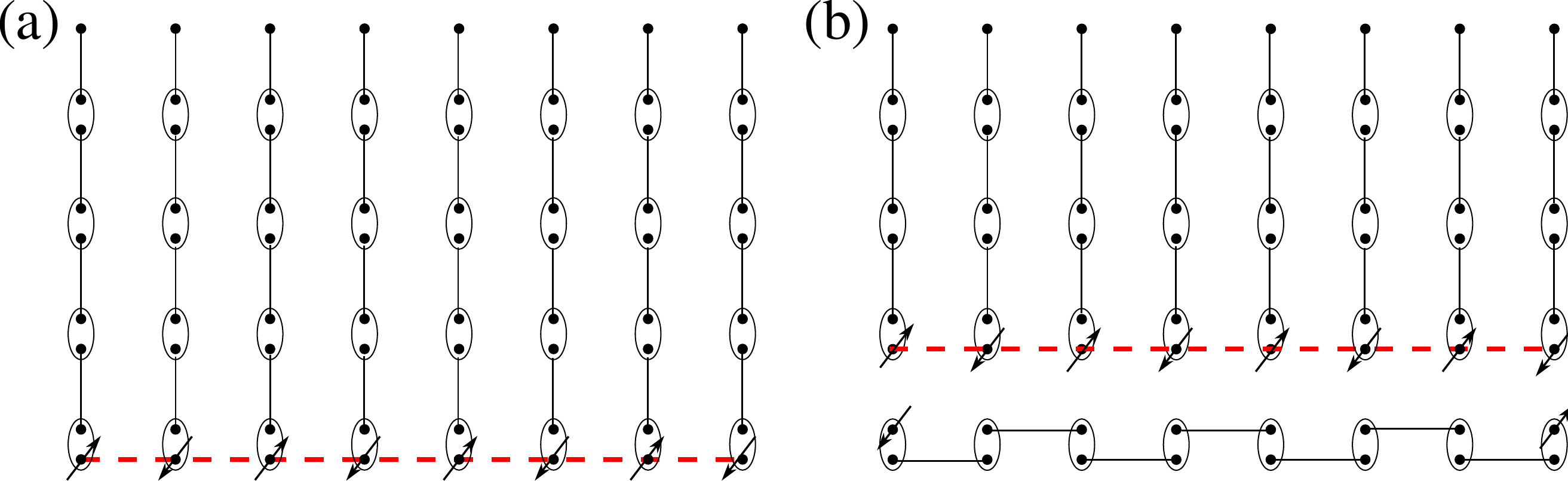}
\caption{(Color online) Valence bond singlet (VBS) sketch of the ground-state in the limit of weak (a) and strong (b) backbone interaction. Each spin-1 (ellipse) is represented by a pair of spin-1/2 (dots), each of which form a VBS singlet with its neighbor so the bulk is in the Haldane  (AKLT) state with single VBS singlet per bond. The unpaired spin-1/2 become an edge state (arrow). In the presence of a non-zero backbone interaction arrows couple into a critical spin-1/2 chain. (a) In the limit when the backbone interaction is weak the critical chain is located on the backbone. (b) For large $J_{bb}>>1$ the system prefers the Haldane state along the backbone and the critical chain is formed out of next sites.}
\label{fig:sketch}
\end{figure}

The two states are sketched in Fig.\ \ref{fig:sketch}, where (a) corresponds to
the limit of weak backbone interaction so the emergent edge states (arrows)
form a critical spin-1/2 chain (dashed red line), and (b) corresponds to the
limit of strong backbone interaction with the Haldane chain on the backbone and
critical spin-1/2 chain on the second sites of the teeth. Interestingly, the
effective size of the critical spin-1/2 chain changes: in the limit of weak and
strong backbone coupling the critical chain is formed out of the edge states of
all $N$ teeth, while in the intermediate coupling the system favors the state
with very long Haldane chain that includes the first tooth, the backbone and
the last tooth, so the critical chain is formed by the second sites on $N-2$
teeth only.
 
We find that the two regimes sketched at the Fig.\ \ref{fig:sketch}(a) and (b)
are connected by a smooth crossover. In both regimes the system
is critical due to the presence of a  critical spin-1/2 chain, regardless of its
location. In the thermodynamic limit the bulk gap is closed for all values of
the backbone interaction. Also, the universality class of the underlying
critical theory in the two regimes is the same. All the arguments above together
with the extremely smooth change of all measured observables (more details
below) suggests that two regimes are indeed connected by a crossover, rather
than a phase transition. 

\begin{figure}[t!]
\includegraphics[width=0.49\textwidth]{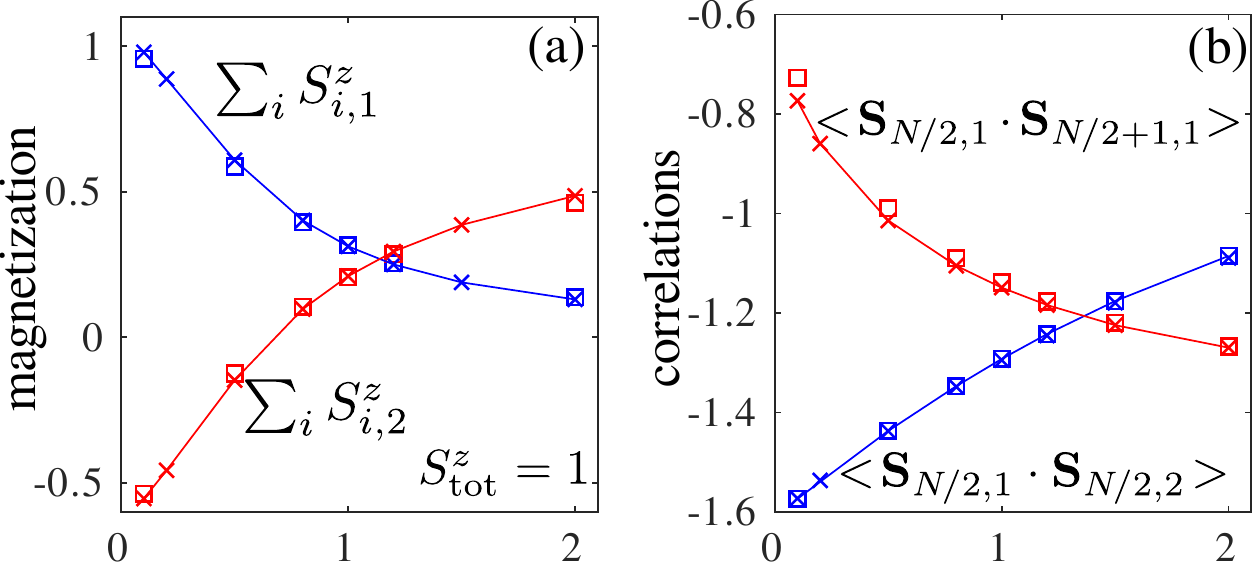}
\caption{(Color online) (a) Total magnetization localized at the backbone (blue) and the second row (red). (b) Local spin-spin correlations on the backbone (red) and between the first two sites on a teeth (blue). In both panels the results are for  square combs  of size  $10\times 10$ (squares) and $20\times 20$ (crosses). A crossover is associated with the crossing point. }
\label{fig:scrossover}
\end{figure}

In order to approximately locate the crossover we look at the location of magnetic excitations. For that we compute the lowest energy state in the
sector of total magnetization $S^z_\mathrm{tot}=1$ and calculate the total
local magnetization along the backbone and along the chain next to the
backbone. The results are summarized in Fig.\ \ref{fig:scrossover}(a). When the
backbone interaction is very small, the backbone has magnetization close to
$\sum_{i=1}^N S^z_{i,1} =1$. The magnetization of the second row is non-zero
due to finite (non-zero) correlation length in the Haldane phase on a teeth; it
takes the opposite sign to the magnetization of the backbone due to strong
antiferromagnetic correlations between the first two sites on the teeth. Upon
increasing the backbone interaction the localization of the magnetic excitation
on the backbone is smeared out and eventually the second row carries higher
magnetization than the backbone. As shown in Fig.\ \ref{fig:scrossover}(a), the two
lines cross when the backbone coupling is about $J_{bb}\approx 1.1$. We performed the
calculations on square combs with $10\times 10$ and $20\times 20$ sites,  finding that the finite-size effects are very small.

As a alternative method, we measure nearest-neighbor correlation in the middle
of the backbone and between the first two sites on the middle tooth. In Fig.\
\ref{fig:cor_graph} we already discussed a qualitative difference between the
local correlation pattern in the two regimes. In Fig.\ \ref{fig:scrossover}(b)
we provide a quantitative comparison. We find that the backbone correlation
becomes stronger than the connect correlations (correlations between the first two sites on a tooth) when the the backbone coupling
exceeds $J_{bb}\approx 1.3$. This value is in a decent agreement with results
on spin-1 localization presented above.

\subsection{Zig-zag backbone and Kosterlitz-Thouless transition}

In this section we show that one can manipulate the emergent critical spin-1/2 chain by adding frustration such as next-nearest-neighbor interaction. In a true
spin-1/2 chain, this would lead to a gapped dimerized phase when the
next-nearest-neighbor coupling constant $J_2$ exceeds $J_2=0.2411$\cite{okamoto}. The
critical phase near $J_2=0$ is separated from the dimerized one by a Kosterlitz-Thouless
(KT) phase transition\cite{Kosterlitz}. Both the critical phase and the KT
transition are characterized by the WZW SU(2)$_1$ universality class. The KT critical line can be identified by vanishing logarithmic corrections,
which are present inside the critical phase due to a marginal operator.  Apart
from the KT transition point, there is another special point, known as the
Majumdar-Ghosh\cite{majum_ghosh} point and located at $J_2=1/2$, where the
ground-state is given by an exactly dimerized state.

\begin{figure}[t!]
\includegraphics[width=0.4\textwidth]{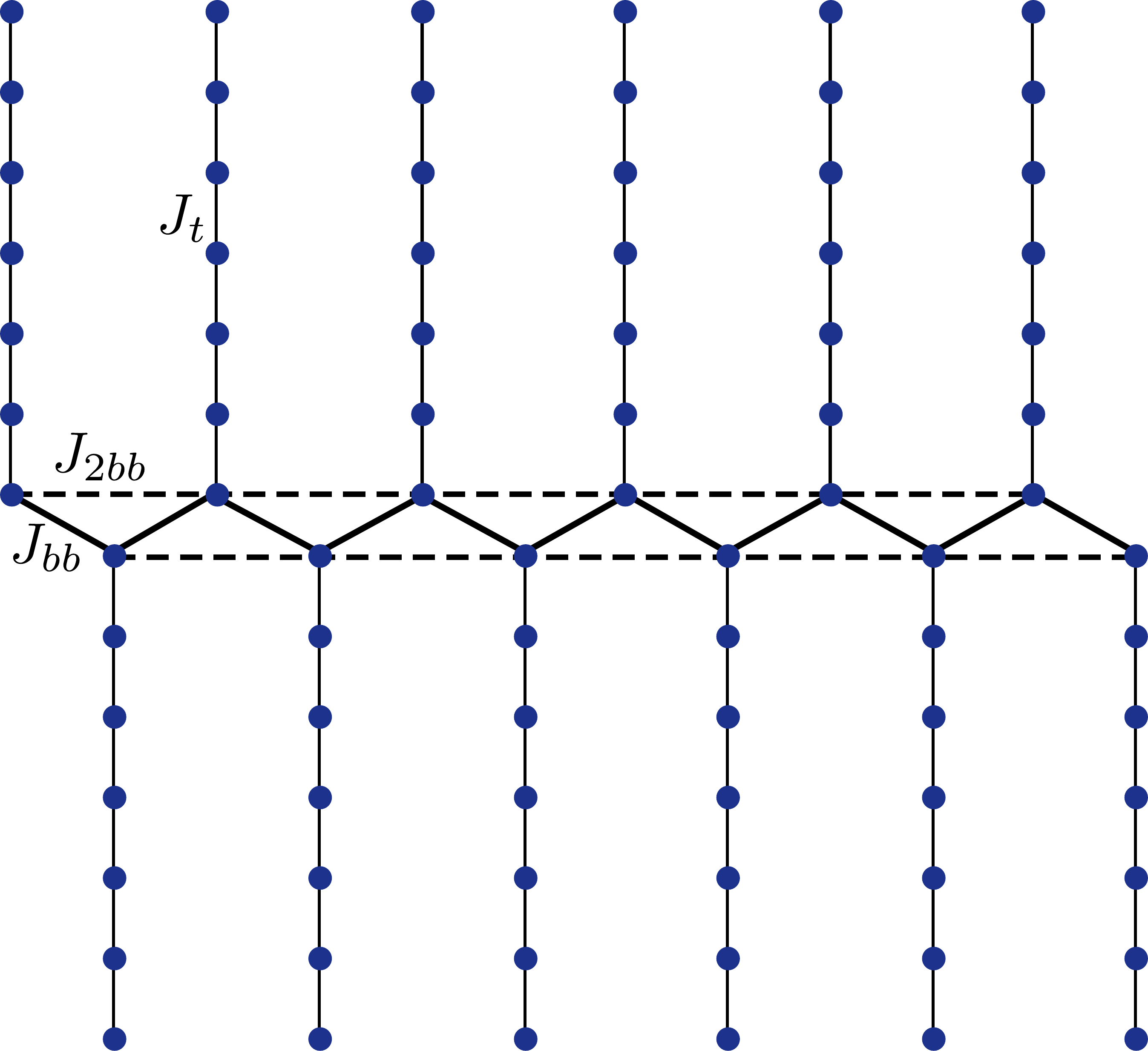}
\caption{(Color online) An alternative representation of a comb lattice, in which both nearest- and next-nearest-neighbor interaction along the backbone naturally appears. 
}
\label{fig:alternative_geometry}
\end{figure}

Here we include an antiferromagnetic next-nearest neighbor interaction along
the backbone of a comb. The most natural way to imagine this is shown in Fig.\
\ref{fig:alternative_geometry} with a zig-zag ladder as a backbone, decorated
with chains/teeth.

The traditional ways to locate the quantum phase transition, such as
finite-size scaling of the order parameter or gap closing, which normally work
well for gapped phases, cannot provide very accurate results for the
Kosterlitz-Thouless transition, because the phase on  one side of the
transition is  critical. Here we closely follow the method used to locate the
end point on the critical lines described in Ref.\onlinecite{dimtrans} and also used
to locate the KT transition in the spin-3/2 chain \cite{chepigas15,chepiga_thesis}.
We briefly review the main idea here. From a CFT point of view, the transition
between the critical and the dimerized phase is driven by tuning the coupling
constant of the marginal operator in the Hamiltonian. When this coupling
constant is negative (up to a convention), it can be renormalized to zero and
the system can be described by WZW SU(2)$_1$ critical theory with non-vanishing
logarithmic corrections that appear due to the renormalization process. By
contrast, when the coupling constant of the marginal operator is positive, it
cannot be renormalized to zero and lead to the gapped dimerized phase. At the
KT transition this coupling constant is equal to zero and therefore the system
is in the WZW SU(2)$_1$ universality class without any logarithmic corrections.
So, the identification of the KT point is equivalent to the identification of
the point where the critical behavior is maximally close to the WZW SU(2)$_1$
even on a finite-size system. 

\begin{figure}[t!]
\includegraphics[width=0.49\textwidth]{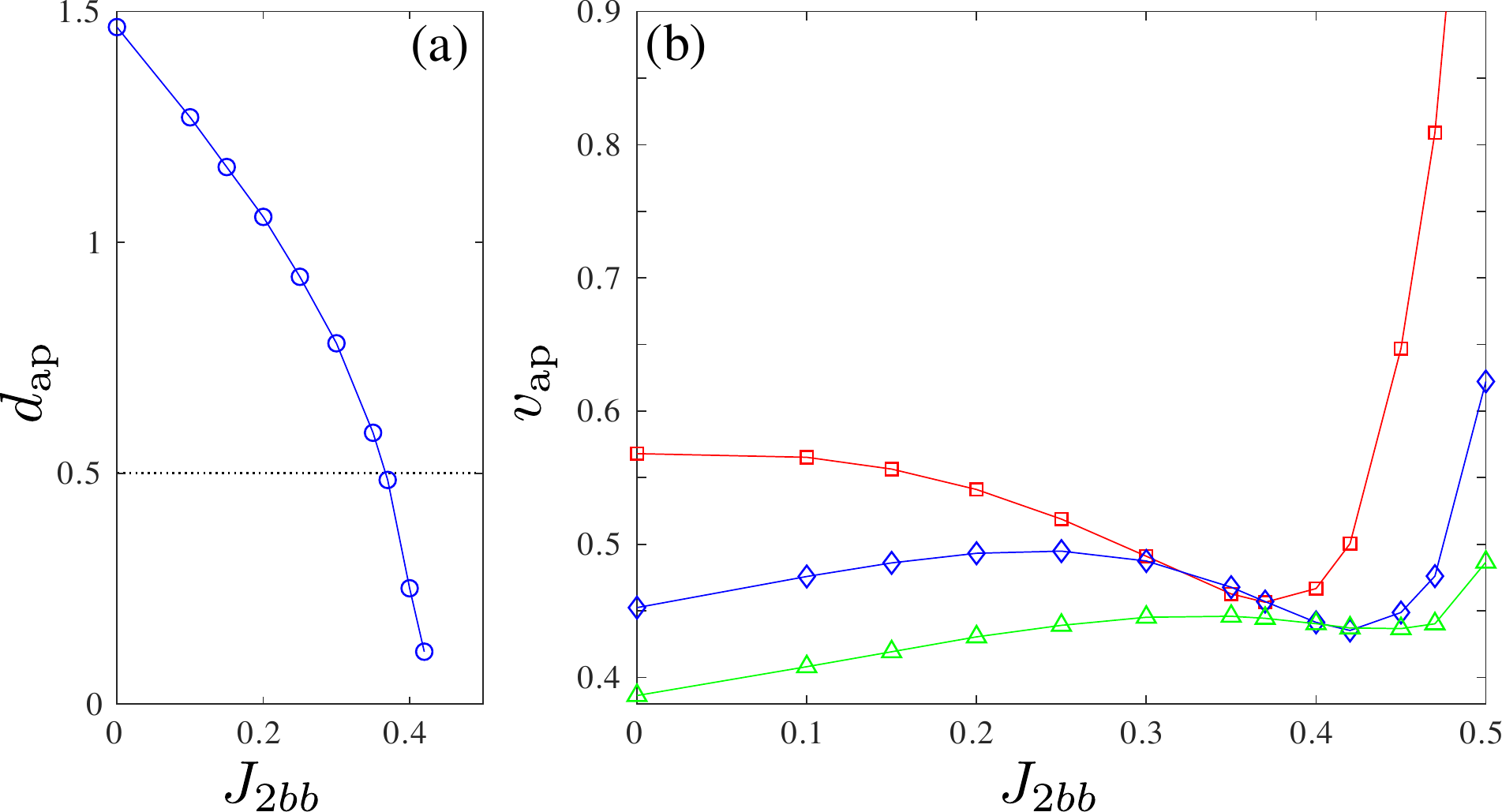}
\caption{(Color online) (a) Apparent critical exponent and (b) apparent velocities as a function of the next-nearest-neighbor coupling along the backbone of a square comb with $20\times 20$ sites and $J_t=J_{bb}=1$. The apparent critical exponent deviates from the CFT prediction $d=1/2$ due to the presence of the logarithmic corrections to the WZW SU(2)$_1$ theory on the left and inside the gapped phase on the right. (b) Velocities extracted for different levels coincide only at the KT point, where the conformal tower is restored. On the left of it the structure of the spectrum is destroyed due to the logarithmic corrections, and on the right, the system is no longer gappless, so the velocities  move away from each other very fast. }
\label{fig:velocity}
\end{figure}

We consider a square comb with $20\times 20$ sites and we set $J_t=J_{bb}=1$.
While tuning the next-nearest-neighbor coupling on the backbone we extract the
apparent critical exponent and velocity. In order to extract the critical
exponent we first note that the ground-state favors dimerization at the edges
of the backbone. This effect is similar to one observed in the simple spin-1/2
chain and remains robust in the presence of decorating teeth. The dimerized
states at the edge of a chain fix the boundary conditions and therefore induce
the Friedel oscillations, which according to boundary conformal field theory
takes the following form:

\begin{equation}
D_{bb}(i,N)\propto\frac{1}{N[\sin(\pi i/N)]^d},
\end{equation}
where $D_{bb}(i,N)$ is an absolute value of the dimerization at site $i$ along the backbone with $N$ teeth. In the absence of logarithmic corrections the critical exponent predicted by boundary CFT is $d=1/2$. When logarithmic corrections are not vanishing, the fit of the Friedel oscillations to this form gives an apparent critical exponent that deviates from the true CFT value. While approaching the thermodynamic limit the deviation slowly (logarithmically) goes to zero and in the thermodynamic limit the critical exponent is equal to $d=1/2$ below and at the KT transition and zero above it. On a finite-size system, the curve is often very smooth. 
The apparent critical exponent extracted from a comb with $20\times 20$ sites is shown in Fig.\ \ref{fig:velocity}(a). One can see that it monotonically decays and crosses the  $d=1/2$ line around $J_{2bb}\approx 0.37$. 

The second way to locate the KT transition is based on the vanishing logarithmic
corrections in the excitation spectra. For critical WZW SU(2)$_1$ spin-1/2
chain with an even number of sites and fixed boundary conditions (see above) the
ground-state is a singlet and the excitation energy scales with the length of
the chain $N$ as

\begin{equation}
  \Delta=\frac{\pi v n}{N},
\end{equation}
where $n=1,4,9$ for the first triplet, quintuplet, and septuplet
correspondingly. Due to logarithmic corrections the scaling can deviate on a
finite-sized systems. One can therefore extract an apparent velocity as
$v_\mathrm{ap}=\frac{\Delta N}{\pi n}$, with the corresponding integer $n$.
When the logarithmic corrections vanish and the structure of the excitation
spectrum  (conformal tower) is preserved on a finite-size system, the apparent
velocities extracted for various levels take the same value. Away from the KT 
transition inside the critical phase the velocities are different,  due to the
presence of the logarithmic corrections, while in the dimerized phase the
spectrum is gapped and its structure changes drastically. Velocities extracted
for three excited states with $S^z_\mathrm{tot}=1,2,3$ as a function of the
next-nearest-neighbor coupling are shown in Fig.\ \ref{fig:velocity}(b). All
velocities come close to each other around $J_{2bb}\approx0.37$, which agrees
with the results from the analysis of the apparent critical
exponent in panel (a).

\begin{figure}[t!]
\includegraphics[width=0.49\textwidth]{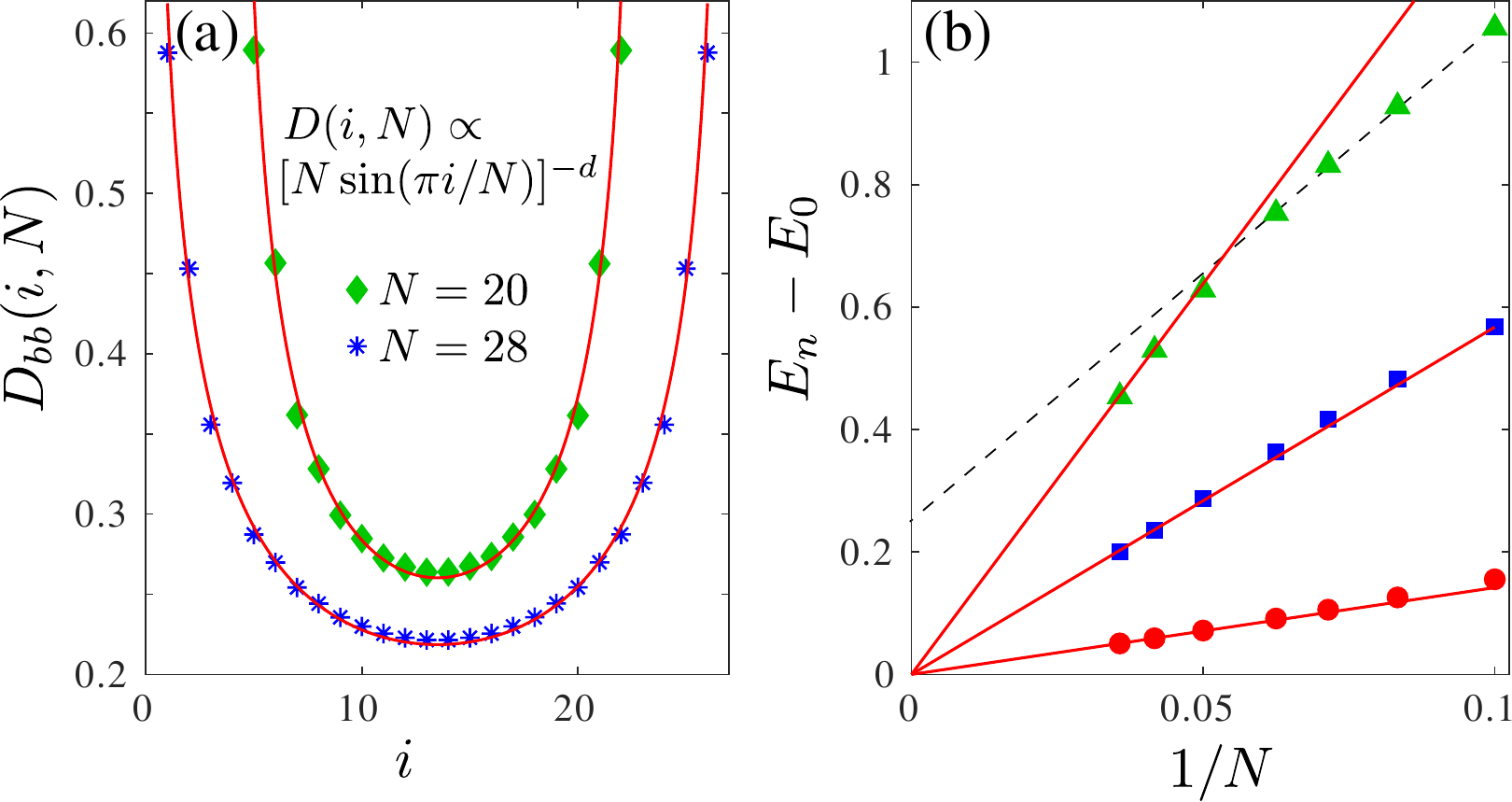}
\caption{(Color online) (a) Dimerization profile (Friedel oscillations) along a backbone of a square comb with $20\times 20$ (green) and $28\times 28$ (blue) sites fit to CFT prediction of Eq. \ref{eq:dimprof}. The resulting critical exponents are $d_{20}\approx 0.485$ and $d_{28}\approx0.483$, in good agreement with the CFT prediction for WZW SU(2)$_1$ $d=1/2$. (b) Finite-size scaling of the excitation energies above the ground-state. Only the lowest levels for each magnetization sectors with $S^z_\mathrm{tot}=1,2,3$ are shown. Symbols are the DMRG data, red lines are the CFT predictions with a reference velocity $v_{ref}=0.451$ (see main text for details). Dashed line is a guide to the eye for the finite-size scaling of a quintuplet state that appears below the corresponding bulk excitation on a small system sizes. }
\label{fig:KT_transition}
\end{figure}

Fig.\ \ref{fig:KT_transition}(a) provides an example of the fit of the Friedel
oscillations on a finite-size comb with $20\times 20$ and $28\times 28$ sites.
For both sizes the extracted critical exponent agrees with the CFT prediction
$d=1/2$ within $4\%$. We also look at the finite-size scaling of the excitation
spectrum (only lowest energy states in the sectors with
$S^z_\mathrm{tot}=0,1,2,3$). The results are summarized in Fig.\
\ref{fig:KT_transition}(b). The reference velocity is estimated by the singlet
triplet gap of the largest available system size ($N=28$, $L=N$) as
$v_\mathrm{ref}=(E_T-E_S)N/(\pi)$. Then we use this non-universal constant to
plot three lines lines of the  CFT prediction $\Delta=(\pi v_\mathrm{ref}n)/N$
for $n=1,4,9$ and see very good agreement with our numerical data. Discrepancy
between the CFT prediction and the data for $S^z_\mathrm{tot}=3$ for
$N^{-1}>0.05$ can be because of low-lying states with a finite gap (dashed
line) which is lower than the first bulk septuplet excitation for system
sizes smaller than $N\approx 20$. 

To summarize, by introducing an additional next-nearest-neighbor frustration
along the backbone we were able to drive the comb into the dimerized state on
the backbone passing through the Kosterlitz-Thouless critical point. For
$J_t=J_{bb}=1$ we locate the KT transition around $J_{2bb}\approx 0.37$, which is
significantly higher than in the simple spin-1/2 chain with $J_1-J_2$
interaction where the transition takes place at $J_2/J_1\approx0.2411$. This
suggest that partial localization of the edge states on the second sites of the
teeth induce an effective ferromagnetic coupling between the
next-nearest-neighbor teeth. This is confirmed by our simple calculation for
$J_{bb}=0.1$ for which the KT transition is located approximately at
$J_{2bb}\approx0.25$, much closer to the KT transition of the original spin-1/2
chain.

\subsection{Higher-order edge states}

\begin{figure}[t!]
\includegraphics[width=0.4\textwidth]{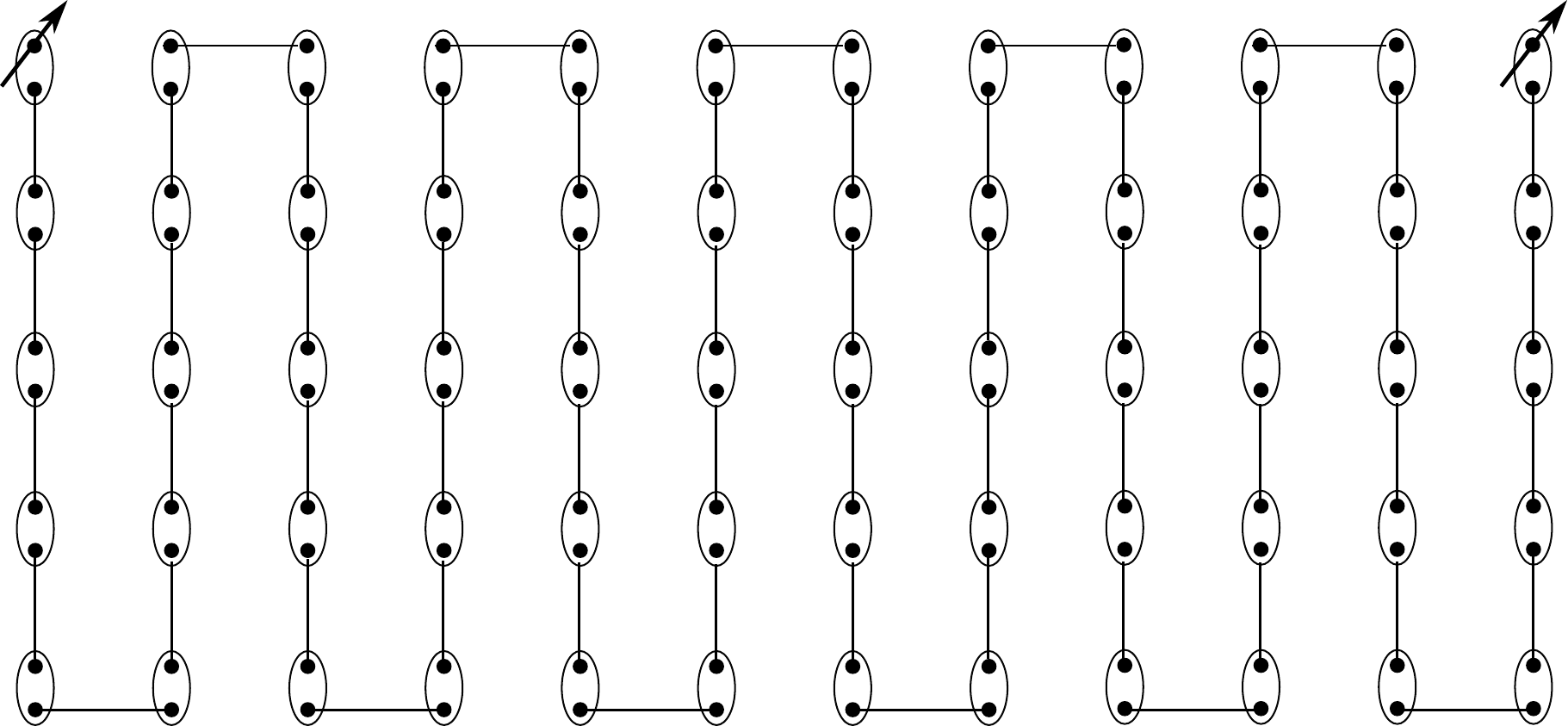}
\caption{ Sketch of the ground-state of the spin-1 Heisenberg comb with with odd number per tooth. Each tooth is in the Haldane phase with spin-1/2 edge states at each end of the tooth. Because the length of the tooth is odd, the edge spin-1/2 form a triplet ground-state and the entire comb is equivalent to a spin-1 chain, the ground-state of which corresponds to the Haldane chain. Since the energy cost of the upper and lower VBS singlets on this effective spin-1 chain are not equal, there is an energetically preferred Haldane state with the fist and the last VBS singlets sitting on the backbone. Therefore the emergent spins-1/2 at the edges of the Haldane chain are localized at the end of the first and the last teeth. }
\label{fig:OddComb}
\end{figure}

We return to our original comb model with all spins-1, including at
the tooth ends.  In a comb with an odd number of sites per teeth, the edge
states of each tooth form a triplet as the ground-state. Therefore each
tooth can be viewed as a composite spin-1 object. By tuning the interaction
along the backbone one also tunes the correlation between the edge spin-1/2 at
the end of the teeth, although the effective interaction between them vanishes
exponentially fast with the length of the teeth.  It is very natural to think
here in terms of the VBS singlets: each tooth with a pair of spin-1/2 edge
states corresponds to a composit spin-1 object. The ground-state is then given
by the Haldane state, where each spin-1/2 is connected by a VBS to one of its
neighbor. In a simple spin-1 chain two spins-1/2 inside a spin-1 object are
perfectly symmetrized, so the Haldane state is not affected by a specific
arrangements of a VBS singlets, as soon as each nearest-neighbor bond contains
one and only one VBS singlet. The situation is a bit different on a comb, since
the interaction between the edge spins-1/2 on a backbone is much stronger than
the interaction between the edge spins at the end of each tooth. As we know,
open boundary condition in the spin-1/2 chain favor dimerization. Most of the interactions between the composite $S=1$'s
 of an effective Haldane chain is due the backbone. So
the total energy of the comb is minimized when the first VBS singlets of the
composite Haldane chain is located at the first bond of the backbone. The next
VBS singlets is them placed between the edge spins-1/2 at the end of teeth 2
and 3, then again on the backbone between teeth 3 and 4 etc. This state is
sketched in Fig.\ \ref{fig:OddComb}.

\begin{figure}[t!]
\includegraphics[width=0.49\textwidth]{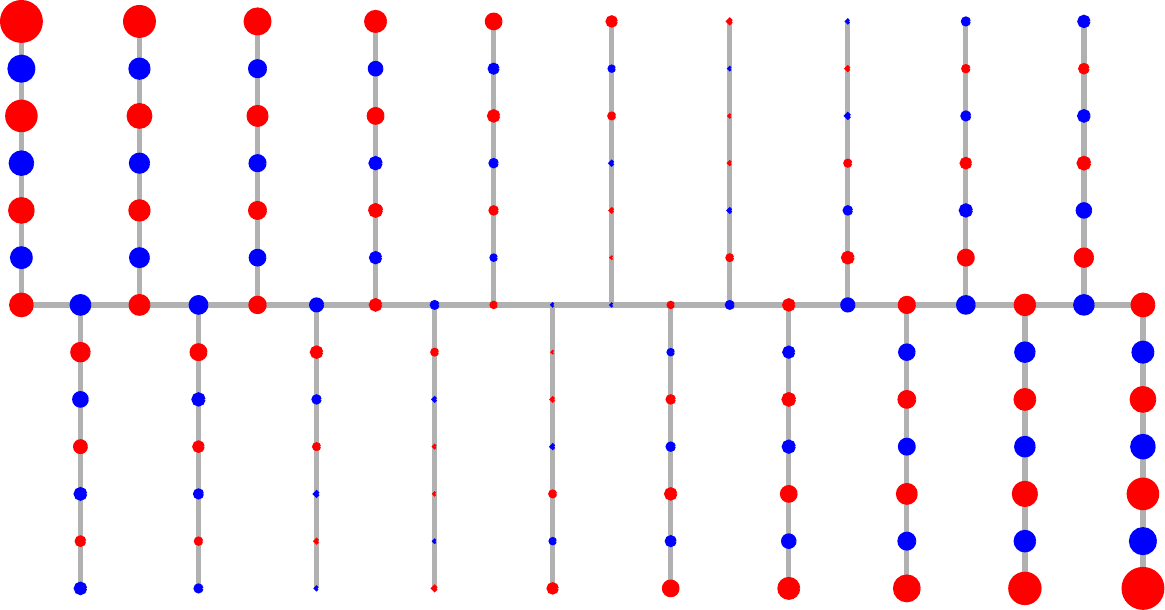}
\caption{(Color online) Local magnetization profile of a spin-1 comb with 20 teeth and 7 spins per tooth, in the sector of total magnetization $S^z_\mathrm{tot}=1$. The edge state of the Haldane state on each tooth form a triplet, so the entire comb is equivalent to the spin-1 chain, which is also in the Haldane phase with one valence bond singlet between the pair of neighboring tooth. The edge states emergent in this Haldane chain are localized at the end of the first and the last teeth - the two furthest points of a comb. On two neighboring teeth the spins equally distant from the backbone are anti-parallel. The maximal polarization is $0.15$}.
\label{fig:edge_states}
\end{figure}

Within the described VBS picture it is clear that the end spins of the first
and last teeth remain unpaired and form the edge states of an entire comb. On
a finite-size system, these two spins-1/2 can couple to each other and form
triplet and singlet in-gap states with an extremely small energy splitting. To
confirm this picture we extract the local magnetization of a comb with
$20\times 7$ sites within the sector of total magnetization
$S^z_\mathrm{tot}=1$, so both edge states are polarized in the same directions.
Fig.\ \ref{fig:edge_states} show the distribution of local magnetization over
the comb. Since the teeth are coupled only along one line - a backbone
- there is some freedom on how to visualize this lattice. In the present case,
we find it instructive to show the results on a two-side comb, but the lattice and the model are equivalent to the one introduced
in Fig.\ \ref{fig:edge_states}.

When plotted as a two-dimensional lattice but  with very special interaction
 these emergent edge states can be viewed as the simplest example of a
higher order edge states discussed in the context of the topological insulators
\cite{benalcazar}.

\begin{figure}[t!]
\includegraphics[width=0.49\textwidth]{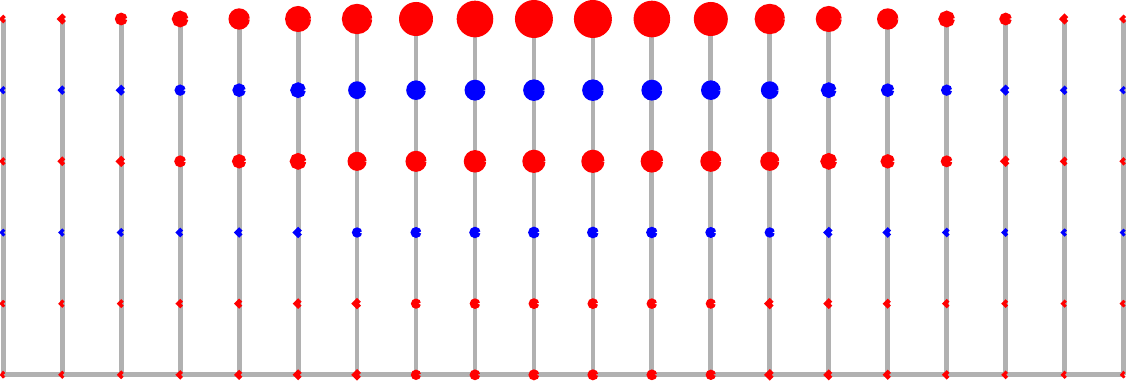}
\caption{(Color online) Local magnetization profile of a spin-1 comb with 20 teeth and 6 spins per tooth, in the sector of total magnetization $S^z_\mathrm{tot}=1$. Each tooth is in the singlet sector and the system is equivalent to a spin-1/2 ladder with weak rungs and one strong and one weak (or absent) leg. The magnetic excitation is delocalized over the whole comb, but most of the weight is on the edge of the teeth. On two neighboring teeth the spins equally distant from the backbone are parallel. The maximal polarization is at the edge of the middle tooth and is equal to $0.08$. }
\label{fig:edge_states_even}
\end{figure}

A comb with even an number of spins per tooth corresponds to a
 spin-1/2 chain decorated with weakly coupled spin-$1/2$'s with antiferromagnetic legs and rungs.  Therefore the
magnetic excitation is mostly localized on a second weak leg, and due to finite
correlation length slightly delocalized along the teeth, as shown in Fig.\
\ref{fig:edge_states_even}.
 
%

\section{Conclusion}
\label{sec:conc}

In the present paper we describe an alternative type of  tree tensor network
applicable to one-dimensional models with complicated on-site clusters. This
numerical set-up partially fills the gap between one-dimensional DMRG and
two-dimensional PEPS (PEPS) tensor network algorithms. While DMRG has
established itself as the most efficient numerical tool for one-dimensional
system of strongly interacting particles, in more complicated cases, such as
a chain of clusters each with substantial entanglement, the
intra and inter-cluster entanglement are both sent through one auxiliary bond,
so the total bond dimension required may be large.
Higher dimensional tensor networks, such as, PEPS, can distribute
entanglement through many links to satisfy the area law naturally, but they
involve complicated algorithms with high complexity in the bond dimension.
The comb tensor network is a useful compromise, with much of the computational
efficiency of DMRG but with the ability to separate intra and inter cluster
entanglement.  Of course, for true two- or higher-dimensional lattices,
the bond dimension within the tooth would grow exponentially, and the comb is
not likely to be useful.

The comb algorithms presented here have great flexibility, including allowing local changes to the Hamiltonian or disorder,
alteration or variability in the local degrees of freedom (e.g. spin value), beyond nearest-neighbor interactions,
the length of the teeth, etc. Along any tooth one has the full flexibility of an MPS along with the favorable
complexity of an MPS, allowing, for example, each tooth to represent a finite higher dimensional or randomly connected cluster.

Of course, the most natural system for a comb tensor network is a system with a comb geometry, or one whose strong local interactions
have the topology of a comb.
In the present paper we
discussed spin-1 Heisenberg model on a comb lattice.
The teeth in this system have a bulk gap and are in a
Haldane phase, which is topological and has emergent edge states. The spin-1/2 edge
states of the teeth are coupled by the backbone interaction 
and form a critical chain. This provides a simple recipe for how
half-integer-spin criticality can be realized with integer spins only.
In a s $S=1/2$ comb, the teeth themselves are critical. The tuning of the backbone interaction also tunes effectively the dimensionality of the entire system. We will present results for this intriguing system elsewhere\cite{chepiga_comb05}.

\section{Acknowledgments}

This work has been supported by the Swiss National Science Foundation and the U.S. National Science Foundation
through DMR-1812558, and the Simons Foundation through the Many-Electron Collaboration.

\bibliographystyle{apsrev4-1}
\bibliography{bibliography}

\end{document}